\documentclass[letterpaper,11pt]{article} 
\pdfoutput=1                             

\usepackage[T1]{fontenc}               

\usepackage{jheppub}                   

\usepackage{mathtools}                 
\usepackage{amssymb,amsfonts}          
\usepackage{mathrsfs}                  
\usepackage{bm}                        

\usepackage{physics}                   

\usepackage{graphicx}                  
\usepackage{subcaption}                

\usepackage{booktabs}                  
\usepackage{tabularx}                  
\usepackage{multirow}                  
\usepackage{makecell}                  
\usepackage[table]{xcolor}             

\usepackage{tikz}                      
\usepackage{tikz-cd}                   
\usetikzlibrary{shapes.geometric,arrows,positioning} 

\usepackage{empheq}                    

\hypersetup{                           
  colorlinks=true,
  linkcolor=blue,
  citecolor=blue,
  urlcolor=blue
}

\newtheorem{remark}{Remark}
\newcommand*\widefbox[1]{\fbox{\hspace{2em}#1\hspace{2em}}}

\newcommand{\nn}{\nonumber}

\title{Two-Mode Janus States: non-Gaussian generalizations of thermofield double
}

\author[a]{Arash Azizi}

\affiliation[a]{{\it The Institute for Quantum Science and Engineering,
Texas A\&M University,\\ College Station, TX 77843, U.S.A.}}

\emailAdd{sazizi@tamu.edu}

\abstract{
We introduce the Two-Mode Janus State (TMJS), a non-Gaussian quantum state defined as a coherent superposition of two distinct Two-Mode Squeezed States (TMSS). This construction serves as a direct, non-Gaussian generalization of the canonical thermofield double (TFD) state, which is itself a single, Gaussian TMSS. We develop a complete analytical framework for the TMJS's arbitrary $k$-th order photon statistics, identifying a new family of "squeezing polynomials" that govern all diagonal and off-diagonal moments. Our central result is that the state's non-Gaussianity is dynamically steerable via an external "Janus phase." This phase acts as a switch, allowing the higher-order coherences ($g^{(k)}$) to be tuned from perfectly thermal (matching the TFD marginal) to deeply sub-Poissonian, a regime marked by strong Wigner negativity. We further establish a physical realization for the TMJS, proposing its generation via coherently superposed Dynamical Casimir Effect (DCE) trajectories, distinguishing it from the static, observer-dependent Unruh effect. The TMJS provides a versatile, interference-enhanced platform for engineering Unruh-DeWitt detector responses and probing non-Gaussian physics in relativistic settings.
}

\begin{document} 
\maketitle
\flushbottom

\section{Introduction}
\label{sec:intro}

A central paradigm in quantum field theory and gravity is the thermofield double (TFD), a pure entangled state constructed on a doubled Hilbert space that, when traced over one-half, yields a perfect thermal state \cite{Israel1976}. Its most celebrated physical realization is the Unruh effect, a cornerstone of quantum field theory in curved spacetime \cite{Unruh1976, Hawking1975, Davies1975, Fulling1973, DeWitt1975PhysicsRep, Birrell_Davies1982}. This effect predicts that a uniformly accelerating observer will perceive the Minkowski vacuum---the state an inertial observer finds to be empty---as a thermal bath with a temperature $T_U = a/2\pi$ proportional to their proper acceleration $a$.

The fundamental origin of this phenomenon is that the Minkowski vacuum, when described in the Rindler basis appropriate for the accelerating observer, is a Two-Mode Squeezed State (TMSS) \cite{UnruhWald1984, Israel1976}. This state perfectly entangles the quantum field modes in, say, the right Rindler wedge, to which the observer is confined, with the modes in the causally disconnected left Rindler wedge \cite{Rindler66}.  The accelerating observer, by definition, can only access the right-wedge. Tracing out the inaccessible left-wedge modes leaves the observer's local field in a precisely thermal state. This Unruh-TMSS is a canonical example of a Gaussian state, as its Wigner function is positive and Gaussian.

Gaussian states, such as the TMSS and coherent states, are foundational resources in quantum information \cite{Weedbrook2012gaussian, Braunstein2005quantum, Adesso2014continuous}. They are readily produced, for example, via parametric down-conversion \cite{Wu1986generation} or four-wave mixing \cite{slusher1985observatio}, and are the basis for continuous-variable (CV) protocols like teleportation and cluster-state computation \cite{Menicucci2006}. However, Gaussian states and operations are insufficient for universal quantum computation, and it is precisely the introduction of non-Gaussianity that provides a key resource for quantum advantage \cite{Ohliger2010limitations, Mari2012Wigner, Albarelli2018Resource}. This establishes a strong motivation to develop controllable, non-Gaussian resources.

To this end, the ``Janus program'' was recently introduced as a framework for engineering such states from coherent superpositions of simple Gaussian components \cite{Azizi2025Janus, Azizi2025Janus_higher, Azizi2025displacedJanus}. The foundational (single-mode) Janus state---a superposition of two squeezed vacua---demonstrates how quantum interference can be engineered to transform the strong bunching ($g^{(k)} \to \infty$) of its components into sub-Poissonian statistics, establishing a universal lower bound of $g^{(2)}(0) = 1/2$ \cite{Azizi2025Janus}. This analysis, which introduced ``squeezing polynomials'' to solve the $k$-th order statistics, was extended to show that a phase-controlled switch can drive all higher-order coherences ($g^{(k \ge 3)}$) to zero \cite{Azizi2025Janus_higher}. The framework also includes displaced Janus (DJ) states, which are distinct from traditional "squeezed-cat" states and provide a complete toolkit for "dialing in" tunable nonclassicality \cite{Azizi2025displacedJanus}.

In this paper, we synthesize these two threads: we use the non-Gaussian Janus framework to generalize the Gaussian Unruh-TFD state. We introduce the \textit{TMJS}, defined as a coherent superposition of two distinct Two-Mode Squeezed States: $|\Psi\rangle = \chi|\xi\rangle + \eta|\zeta\rangle$. This state serves as the direct, non-Gaussian analogue of the Unruh-TFD state. We derive the complete set of $k$-th order factorial moments for the TMJS, providing a full analytical framework for its photon statistics. Our analysis identifies the two-mode family of "squeezing polynomials" $P_k(x)$ (Eq.~\ref{eq:Fk_final_poly}) that govern the diagonal and off-diagonal kernels for all higher-order coherences.

The central result is the demonstration of a "Janus switch" for both single-mode and cross-mode statistics. By tuning the external Janus phase ($\delta$) and the relative squeezing phase ($\Delta$), the statistics of the single-mode marginal---the part accessible to an accelerated observer---can be steered from perfectly thermal ($g_a^{(k)}=k!$) to a regime of deep, coherent suppression (Figs.~\ref{fig:single_mode_g2_landscape}–\ref{fig:single_mode_g4_landscape}). Simultaneously, the characteristic divergence of cross-mode correlations ($g_{ab}^{(k)}$) in a single TMSS is inverted into a suppression of many orders of magnitude (Fig.~\ref{fig:janus_coherence_switch}). We show this suppression is a definitive signature of the state's non-Gaussianity, which manifests as Wigner negativity in phase space (Fig.~\ref{fig:janus_wigner_gallery}) \cite{Kenfack2004}. 

The remainder of this paper is organized as follows.
In Sec.~\ref{sec:definitions}, we establish our notation and review the definitions of the standard two-mode squeezed state (TMSS) and the TMJS.
In Sec.~\ref{sec:derivations}, we present the core calculations of the paper. We first derive the arbitrary $k$-th order factorial moments for a single TMSS, identifying the family of squeezing polynomials that govern them. We then use these results as the diagonal kernels to derive the full, phase-steerable coherence functions for the TMJS.
In Sec.~\ref{sec:vis}, we visualize and analyze these coherence functions. We present coherence landscapes that demonstrate the "Janus switch" for both single-mode and cross-mode statistics, and connect this behavior to the non-Gaussian, negative regions of the Wigner function.
In Sec.~\ref{sec:physical_realization}, we discuss the physical realization of this state, clarifying the distinction between the observer-dependent Unruh effect and state-generating physical processes like the Dynamical Casimir Effect (DCE) \cite{Moore1970, Fulling_Davies1976, Nation2012Nori}. We show how coherently superposed mirror trajectories can, in principle, generate a TMJS.
We offer our conclusions in Sec.~\ref{sec:conclusion}.
Finally, in Appendix~\ref{app:su11_proof}, we provide a detailed proof of the SU(1,1) disentangling theorem used in our derivations.

\section{Formalism}
\label{sec:definitions}

\subsection{The Two-Mode Squeezed State (TMSS)}
\label{subsec:tmss_def}

This section fixes the notation and collects the elementary tools used throughout the paper. We begin with a system of two bosonic modes, described by operators \(a\) and \(b\), which satisfy the canonical commutation relations \([a,a^\dagger]=[b,b^\dagger]=1\), with all other commutators vanishing. The fundamental operator for generating entanglement between these modes is the two–mode squeezing operator, defined as
\begin{align}
S_2(\xi)\equiv \exp\big\{\xi\,a^\dagger b^\dagger-\xi^*\,ab\big\},
\qquad
\xi=r\,e^{i\theta},\quad r\ge 0.
\label{eq:tmss_def}
\end{align}
Here, $r \ge 0$ is the squeezing parameter and $\theta$ is the squeezing phase.

To analyze the properties of this operator, it is convenient to introduce the standard generators of the SU(1,1) algebra:
\begin{align}
K_+\equiv a^\dagger b^\dagger,\qquad
K_-\equiv ab,\qquad
K_0\equiv \tfrac{1}{2}\big(a^\dagger a+b^\dagger b+1\big).
\end{align}
These generators obey the commutation relations \([K_0,K_\pm]=\pm K_\pm\) and \([K_+,K_-]=-2K_0\). Using these relations, the unitary squeezing operator can be re-written in a factorized or "disentangled" form. A standard operator identity for SU(1,1) (a proof of which is provided in Appendix~\ref{app:su11_proof}) gives:
\begin{align}
S_2(\xi)
= \exp\big\{\alpha K_+\big\}\,
(1-|\alpha|^2)^{K_0}\,
\exp\big\{-\alpha^* K_-\big\},
\qquad \alpha\equiv \tanh r\,e^{i\theta}.
\label{eq:su11_disentangle}
\end{align}
A crucial application of this theorem is its action on the two-mode vacuum state \(|0,0\rangle\). Since the vacuum is annihilated by the lowering operator \(K_-\) (i.e., $K_-|0,0\rangle = 0$) and is an eigenstate of \(K_0\) with eigenvalue \(1/2\), the action of \eqref{eq:su11_disentangle} simplifies significantly to the useful normal–ordered form:
\begin{align}
S_2(\xi)\,|0,0\rangle
&=\exp\big\{\alpha K_+\big\}\,
(1-|\alpha|^2)^{K_0}\,
\exp\big\{-\alpha^* K_-\big\}\,|0,0\rangle \nonumber \\
&=\exp\big\{\alpha K_+\big\}\,
(1-|\alpha|^2)^{K_0}\,|0,0\rangle \nonumber \\
&=\sqrt{1-|\alpha|^2}\;\exp\{\alpha\,a^\dagger b^\dagger\}|0,0\rangle,
\qquad \alpha=\tanh r\,e^{i\theta}.
\label{eq:vacuum_action_closed}
\end{align}
Expanding the exponential in this compact expression produces the familiar Fock–space expansion of the resulting state, known as the two–mode squeezed state (TMSS). We also define the shorthand parameters $x$ (based on $\tanh^2 r$) and $\bar n_r$ (the mean photon number):
\begin{align}
|\xi\rangle
=\frac{1}{\cosh r}\sum_{n=0}^{\infty}\big(\tanh r\,e^{i\theta}\big)^{n}|n,n\rangle,
\qquad
x\equiv\tanh^{2}r,\qquad
\bar n_r\equiv\sinh^{2}r=\frac{x}{1-x}.
\label{eq:tmss_fock}
\end{align}

Alternatively, the key identity \eqref{eq:vacuum_action_closed} can be derived from a more fundamental first principle: the uniqueness of states defined by annihilation conditions \cite{Azizi2025Unitary_TFD}. That argument proceeds in two steps. First, one proves that the unique normalizable state $|\psi\rangle$ satisfying the coupled annihilation conditions $(a-\alpha b^{\dagger})|\psi\rangle=0$ and $(b-\alpha a^{\dagger})|\psi\rangle=0$ must be the twin-Fock superposition $|\psi\rangle=\sqrt{1-|\alpha|^{2}}\sum_{m=0}^{\infty}\alpha^{m}|m,m\rangle$ \cite{Azizi2025Unitary_TFD}. Second, one shows (e.g., using the Baker-Campbell-Hausdorff formula) that the unitary state $S_2(\xi)|0,0\rangle$ also satisfies these exact annihilation conditions, with the identification $\alpha = \tanh r e^{i\theta}$ \cite{Azizi2025Unitary_TFD}. Because the solution is unique, the two states must be identical, thus rigorously proving Eq. \eqref{eq:vacuum_action_closed} without recourse to the disentangling theorem.

\subsection{The Two-Mode Janus State (TMJS)}
\label{subsec:TMJS}

In Refs.~\cite{Azizi2025Janus, Azizi2025Janus_higher, Azizi2025displacedJanus} we introduced the \emph{Janus state} as a coherent superposition of two single-mode squeezed vacua. Let
\begin{align}
S_1(\xi)\equiv \exp\!\Big[\tfrac12\big(\xi\,a^{\dagger 2}-\xi^{\ast}a^{2}\big)\Big],\qquad
|\xi\rangle_1=S_1(\xi)\,|0\rangle,\qquad
\xi=r\,e^{i\theta},
\end{align}
and analogously \( |\zeta\rangle_1=S_1(\zeta)|0\rangle \) with \( \zeta=s\,e^{i\phi} \).
The single-mode Janus state was defined as
\begin{align}
|{\Psi}\rangle_{1}\;=\;\chi\,|\xi\rangle_{1}\;+\;\eta\,e^{\,i\delta}\,|\zeta\rangle_{1},
\label{eq:single_mode_janus_def}
\end{align}
with real amplitudes \( \chi=|\chi| \ge 0 \), \( \eta=|\eta| \ge 0 \), and an external \emph{Janus phase} \( \delta \) that steers interference in moments and cumulants.\smallskip

Now we introduce TMJS by extending the Janus philosophy from single-mode squeezing to the \emph{two-mode} Gaussian resource generated by \(S_2\) across modes \(a\) and \(b\). Introduce a second two-mode squeezed vacuum,
\begin{align}
|\zeta\rangle=\frac{1}{\cosh s}\sum_{m=0}^{\infty}\big(\tanh s\,e^{i\phi}\big)^{m}|m,m\rangle,
\qquad
y\equiv\tanh^{2}s,\qquad
\bar n_s\equiv\sinh^{2}s=\frac{y}{1-y}.
\label{eq:second_tmss}
\end{align}
The \emph{TMJS} is the coherent superposition
\begin{align}
|\Psi\rangle=\chi\,|\xi\rangle+\eta\,|\zeta\rangle,
\qquad
\chi=|\chi|\in\mathbb{R}_{\ge 0},\quad
\eta=|\eta|\,e^{i\delta},
\label{eq:tmjs_state_pre}
\end{align}
where \( |\xi\rangle \) and \( |\zeta\rangle \) are two distinct TMSSs (generally different \(\{r,\theta\}\) and \(\{s,\phi\}\)). For any normally ordered observable \(\hat O\), expectation values obey the master bilinear rule
\begin{align}
\langle \hat O\rangle_\Psi
=|\chi|^2\langle\xi|\hat O|\xi\rangle
+|\eta|^2\langle\zeta|\hat O|\zeta\rangle
+2|\chi||\eta|\,\Re\!\big[e^{-i\delta}\langle\zeta|\hat O|\xi\rangle\big].
\label{eq:bilinear_rule_pre}
\end{align}
Interference is governed by the complex overlap parameter
\begin{align}
z\equiv e^{i(\theta-\phi)}\,\tanh r\,\tanh s,
\qquad |z|<1,
\label{eq:z_def}
\end{align}
which serves as the argument of all cross-state kernels. Closed-form expressions for
\(\langle \zeta|(a^\dagger)^k a^k|\xi\rangle\) and
\(\langle \zeta|(a^\dagger b^\dagger)^k(ab)^k|\xi\rangle\),
together with the squeezing polynomials \(P_k(u)\) that control higher-order coherences, are derived in Sec.~\ref{sec:derivations}.

\section{Coherence, Statistics, and Results}
\label{sec:derivations}

\subsection{Coherence of a Single TMSS (Diagonal Kernels)}
\label{subsec:tmss_coherence}

We now derive the factorial moments for a single TMSS, $|\xi\rangle$, using the parameters defined in Eq. \eqref{eq:tmss_fock}. These results will serve as the "diagonal" terms in the Janus state calculation.

\subsubsection{Cross-mode coherence and Squeezing Polynomials}

Our first goal is to compute the cross-mode \(k\)-th factorial moment, \(\langle\xi|(a^\dagger b^\dagger)^k(ab)^k|\xi\rangle\), for any integer \(k\ge 0\) and \(|x|<1\). This calculation is most directly performed in the Fock basis. The action of the lowering operator \((ab)^k\) on a twin-Fock state $|n,n\rangle$ produces the falling factorial:
\begin{align}
(ab)^k|n,n\rangle
= n(n{-}1)\cdots(n{-}k{+}1)\,|n{-}k,n{-}k\rangle .
\end{align}
By taking the adjoint and using the orthonormality of the Fock basis, the expectation value becomes a sum over $n$. Using the parameter $x=\tanh^2 r$ and $\cosh^{-2} r = 1-x$, this sum is
\begin{align}
\langle\xi|(a^\dagger b^\dagger)^k(ab)^k|\xi\rangle
= \frac{1}{\cosh^{2}r}\sum_{n=0}^{\infty}\Big[n(n{-}1)\cdots(n{-}k{+}1)\Big]^2\,(\tanh^{2}r)^{n}
= (1-x)\sum_{n=0}^{\infty}\big[n^{\underline{k}}\big]^2 x^{n},
\label{eq:Fk_series}
\end{align}
where \(n^{\underline{k}}=n(n-1)\cdots(n-k+1)\).

To evaluate this sum, we use a generating function identity based on the differential operator \(x^k \partial_x^k\), which acts on $x^n$ to produce the falling factorial:
\begin{align}
x^{k}\,\partial_{x}^{k}\,x^{n}=n^{\underline{k}}\,x^{n}\qquad (n,k\in\mathbb{N}).
\label{eq:xkdk_identity}
\end{align}
Since our sum \eqref{eq:Fk_series} involves the square of the falling factorial, \([n^{\underline{k}}]^2\), we can express it by applying this differential operator twice to the basic geometric series \(\sum x^n = (1-x)^{-1}\):
\begin{align}
\sum_{n=0}^{\infty}\big[n^{\underline{k}}\big]^2 x^{n}
= x^{k}\partial_{x}^{k}\left[x^{k}\partial_{x}^{k}\left(\sum_{n=0}^{\infty}x^{n}\right)\right]
= x^{k}\partial_{x}^{k}\left[x^{k}\partial_{x}^{k}\left(\frac{1}{1-x}\right)\right].
\end{align}
We can now define the full, unnormalized moment $F_k(x)$ by re-inserting the prefactor from \eqref{eq:Fk_series}:
\begin{align}
F_{k}(x)\;\equiv\;\langle\xi|(a^\dagger b^\dagger)^k(ab)^k|\xi\rangle
= (1-x)\,x^{k}\partial_{x}^{k}\left[x^{k}\partial_{x}^{k}\left(\frac{1}{1-x}\right)\right].
\label{eq:Fk_def}
\end{align}
We evaluate this expression from the inside out. The action of the inner operator is
\begin{align}
\partial_{x}^{k}\left(\frac{1}{1-x}\right)=\frac{k!}{(1-x)^{k+1}}
\quad\Rightarrow\quad
x^{k}\partial_{x}^{k}\left(\frac{1}{1-x}\right)=\frac{k!\,x^{k}}{(1-x)^{k+1}} .
\label{eq:inner}
\end{align}
Applying the outer differential operator, \(x^{k}\partial_{x}^{k}\), to this intermediate result requires the use of the Leibniz rule for the $k$-th derivative of a product:
\begin{align}
\partial_{x}^{k}\big[x^{k}(1-x)^{-(k+1)}\big]
&=\sum_{j=0}^{k}\binom{k}{j}\,\big[\partial_{x}^{k-j}x^{k}\big]\,\big[\partial_{x}^{j}(1-x)^{-(k+1)}\big] \nonumber\\
&=\sum_{j=0}^{k}\binom{k}{j}\,\left[\frac{k!}{j!}\,x^{\,j}\right]\,\left[\frac{(k+j)!}{k!}\,(1-x)^{-(k+1+j)}\right] \nonumber\\
&=\sum_{j=0}^{k}\binom{k}{j}\,\frac{(k+j)!}{j!}\,x^{\,j}\,(1-x)^{-(k+1+j)}.
\end{align}
To complete the calculation of $F_k(x)$, we multiply this sum by the remaining factors from \eqref{eq:Fk_def}, which are the leading \(x^{k}\) and the prefactor \((1-x)\). This combination yields the final rational form:
\begin{align}
F_{k}(x)
&=\frac{(k!)^{2}\,x^{k}}{(1-x)^{2k}}
\sum_{j=0}^{k}\binom{k}{j}\binom{k{+}j}{j}\,x^{\,j}(1-x)^{\,k-j}.
\label{eq:Fk_binomial_weight}
\end{align}
This finite sum can be written compactly using a hypergeometric representation:
\begin{align}
\sum_{j=0}^{k}\binom{k}{j}\binom{k{+}j}{j}\,x^{\,j}(1-x)^{\,k-j}
=(1-x)^{k}\;{}_2F_{1}\left(-k,\,k{+}1;\,1;\,\frac{-x}{1-x}\right)
={}_2F_{1}(-k,-k;1;x),
\label{eq:2F1_equality}
\end{align}
where the last equality follows directly from Euler’s transformation,
\(
{}_2F_1(a,b;c;z)=(1-z)^{-a}\,{}_2F_1\big(a,\,c-b;\,c;\,\tfrac{z}{z-1}\big)
\),
by setting \(a=-k\), \(b=k+1\), \(c=1\), and \(z=-x/(1-x)\).

We can therefore express $F_k(x)$ as a ratio involving a new polynomial family. It is conventional and highly useful to separate this expression into its polynomial part and its singular part (which diverges as $x\to 1$). We therefore define the \emph{squeezing polynomial} family, $P_k(x)$, as the polynomial numerator:
\begin{align}
F_{k}(x)=\frac{P_{k}(x)}{(1-x)^{2k}},
\qquad
P_{k}(x)=(k!)^{2}\,x^{k}\,{}_2F_{1}(-k,-k;1;x)
=(k!)^{2}\,x^{k}\sum_{j=0}^{k}\binom{k}{j}^{2}x^{\,j}.
\label{eq:Fk_final_poly}
\end{align}
Equivalently, this polynomial is related to the standard Legendre polynomial \(\mathrm{P}_{k}\) by
\begin{align}
P_{k}(x)=(k!)^{2}\,x^{k}(1-x)^{k}\,\mathrm{P}_{k}\left(\frac{1+x}{1-x}\right).
\end{align}
Before deriving the cross-mode coherence, we first apply these results to the simpler case of the single-mode factorial moments.

\subsubsection{Single-mode coherence: thermal statistics}
The $k$-th order coherence function for a single mode is formally defined as the $k$-th factorial moment normalized by the $k$-th power of the mean photon number:
\begin{align}
g^{(k)}_{a}(\xi)\equiv \frac{\langle (a^\dagger)^k a^k\rangle_\xi}{\langle a^\dagger a\rangle_\xi^k}.
\end{align}
To compute the numerator $\langle (a^\dagger)^k a^k\rangle_\xi$, we trace over the reduced density matrix of mode $a$. This state is thermal, $\rho_a = (1-x) \sum_n x^n |n\rangle\langle n|$, so the expectation value is
\begin{align}
\langle (a^\dagger)^k a^k\rangle_\xi
= \text{Tr}\left(\rho_a a^{\dagger k} a^k\right)
= (1-x) \sum_{n=0}^\infty x^n \langle n | a^{\dagger k} a^k | n \rangle
= (1-x) \sum_{n=0}^\infty n^{\underline{k}} x^n.
\end{align}
This sum is a standard generating function, which can be found by applying the differential operator from \eqref{eq:xkdk_identity} to the geometric series:
\begin{align}
\sum_{n=0}^{\infty}n^{\underline{k}}\,x^{n} 
= \sum_{n=0}^{\infty} x^k \partial_x^k (x^n) 
= x^k \partial_x^k \left( \sum_{n=0}^\infty x^n \right) 
= x^{k}\partial_{x}^{k}\left(\frac{1}{1-x}\right)
= \frac{k!\,x^{k}}{(1-x)^{k+1}}.
\end{align}
Substituting this result into our expression for the $k$-th moment gives:
\begin{align}
\langle (a^\dagger)^k a^k\rangle_\xi
=(1-x) \left[ \frac{k!\,x^{k}}{(1-x)^{k+1}} \right]
= \frac{k!\,x^{k}}{(1-x)^{k}}.
\end{align}
The mean photon number $\bar n_r = \langle a^\dagger a \rangle_\xi$ is the $k=1$ case, $\bar n_r = x/(1-x)$, which matches our definition in Eq.~\eqref{eq:tmss_fock}. We can therefore write the $k$-th factorial moment compactly in terms of the mean photon number:
\begin{align}
\langle (a^\dagger)^k a^k\rangle_\xi = k! \left( \frac{x}{1-x} \right)^k = k!\,\bar n_r^{k}.
\end{align}
Finally, we assemble the normalized coherence function $g^{(k)}_a(\xi)$:
\begin{align}
g^{(k)}_{a}(\xi) = \frac{\langle (a^\dagger)^k a^k\rangle_\xi}{\langle a^\dagger a \rangle_\xi^k} = \frac{k!\,\bar n_r^{k}}{(\bar n_r)^k} = k!.
\label{eq:gka_tmss}
\end{align}
This result, $g^{(k)}_a = k!$, confirms the well-known fact that the single-mode statistics of a pure two-mode squeezed vacuum, when one mode is traced out, are perfectly thermal.

\subsubsection{Cross-mode coherence and final checks}

We now return to the cross-mode coherence, $g^{(k)}_{ab}(\xi)$. This is obtained by normalizing $F_k(x)$ from Eq.~\eqref{eq:Fk_final_poly} by the $k$-th power of the product of the mean photon numbers, \(\langle a^\dagger a\rangle_\xi\,\langle b^\dagger b\rangle_\xi = \bar n_r^2 = (x/(1-x))^2\). This gives the final, compact form for the coherence:
\begin{empheq}[box=\widefbox]{align}
g^{(k)}_{ab}(\xi)
=\frac{F_{k}(x)}{\big(\langle a^\dagger a\rangle_\xi\,\langle b^\dagger b\rangle_\xi\big)^{k}}
=\frac{P_{k}(x)/(1-x)^{2k}}{\left(x/(1-x)\right)^{2k}}
=\frac{P_{k}(x)}{x^{2k}}
=(k!)^{2}\,x^{-k}\sum_{j=0}^{k}\binom{k}{j}^{2}x^{\,j}
\label{eq:gkab_tmss}
\end{empheq}
As final checks, we can evaluate this for low values of $k$ using the polynomial $P_k(x)$ from Eq.~\eqref{eq:Fk_final_poly}.

\noindent For \(k=1\), \eqref{eq:Fk_final_poly} gives \(P_{1}(x)=x(1+x)\). Substituting into \eqref{eq:gkab_tmss} yields:
\begin{align}
g^{(1)}_{ab}(\xi) = \frac{P_1(x)}{x^2} = \frac{x(1+x)}{x^2} = \frac{1+x}{x} = 1 + \frac{1}{x} = 1 + \coth^2 r = 2+\sinh^{-2}r.
\end{align}
For \(k=2\), \eqref{eq:Fk_final_poly} gives \(P_{2}(x)=4x^{2}(1+4x+x^{2})\), which gives:
\begin{align}
g^{(2)}_{ab}(\xi) = \frac{P_2(x)}{x^4} = \frac{4x^{2}(1+4x+x^{2})}{x^4} = \frac{4(1+4x+x^2)}{x^2}.
\end{align}
For \(k=3\), \eqref{eq:Fk_final_poly} gives \(P_{3}(x)=36x^{3}(1+9x+9x^2+x^3)\), which gives:
\begin{align}
g^{(3)}_{ab}(\xi) = \frac{P_3(x)}{x^6} = \frac{36(1+9x+9x^2+x^3)}{x^3}.
\end{align}
For \(k=4\), \eqref{eq:Fk_final_poly} gives \(P_{4}(x)=576x^{4}(1+16x+36x^2+16x^3+x^4)\), which gives:
\begin{align}
g^{(4)}_{ab}(\xi) = \frac{P_4(x)}{x^8} = \frac{576(1+16x+36x^2+16x^3+x^4)}{x^4}.
\end{align}

\subsection{\texorpdfstring{Arbitrary $k$-th order coherences for TMJS}{}}
\label{subsec:tmjs_coherence}

We now construct the Two–Mode Janus State (TMJS) by forming a coherent superposition of the two TMSSs, $|\xi\rangle$ and $|\zeta\rangle$, using the parameters defined in Sec.~\ref{sec:definitions}. The state is defined as
\begin{align}
|\Psi\rangle=\chi\,|\xi\rangle+\eta\,|\zeta\rangle,
\qquad
\chi=|\chi|\in\mathbb{R}_{\ge 0},\quad
\eta=|\eta|\,e^{i\delta},
\end{align}
where \(\xi=r\,e^{i\theta}\) and \(\zeta=s\,e^{i\phi}\) are the complex squeeze parameters of the two constituent states, and $\delta$ is the external "Janus" phase governing their interference. The expectation value of any normally ordered observable \(\hat O\) in the TMJS state is given by the fundamental bilinear expansion rule:
\begin{align}
\langle\hat O\rangle_\Psi
=|\chi|^2\langle\xi|\hat O|\xi\rangle
+|\eta|^2\langle\zeta|\hat O|\zeta\rangle
+2|\chi||\eta|\,\Re\big[e^{-i\delta}\,\langle\zeta|\hat O|\xi\rangle\big].
\label{eq:tmjs_master_bilinear}
\end{align}
This expression cleanly separates the expectation value into a classical mixture of the two states (the first two terms) and a quantum interference term (the final term). Note that the interference phase \(\delta\) always appears with a minus sign inside the real part operation, \(\Re[\cdot]\).

To compute the $k$-th order moments, we need the diagonal kernels derived in the preceding subsection, and the off-diagonal (cross-state) kernels. We can derive these off-diagonal kernels using the exact same generating function method as in Sec.~\ref{subsec:tmss_coherence}. The mathematical structure of the calculation is identical, with the only change being the parameters that form the underlying geometric series. Let's calculate the cross-mode kernel $\langle \zeta | (a^\dagger b^\dagger)^k (ab)^k | \xi \rangle$. We have:
\begin{align}
\langle \zeta | (a^\dagger b^\dagger)^k (ab)^k | \xi \rangle
&= \sum_{m,n} \frac{(\tanh s e^{-i\phi})^m}{\cosh s} \frac{(\tanh r e^{i\theta})^n}{\cosh r}
\langle m,m | (a^\dagger b^\dagger)^k (ab)^k | n,n \rangle \nn \\
&= \frac{1}{\cosh r \cosh s} \sum_n [n^{\underline{k}}]^2 \big( \tanh r \tanh s\, e^{i(\theta-\phi)} \big)^n \nn \\
&= \frac{1}{\cosh r \cosh s} \sum_{n=0}^\infty [n^{\underline{k}}]^2 z^n.
\end{align}
In the first step, we expanded both states in the Fock basis. In the second, we used the fact that the operator $\langle m,m | \dots | n,n \rangle$ is number-preserving and thus non-zero only for $m=n$, collapsing the double sum. In the third, we used $\langle n,n | (a^\dagger b^\dagger)^k (ab)^k | n,n \rangle = [n^{\underline{k}}]^2$. In the final step, we substituted the definition of $z$ from Eq.~\eqref{eq:z_def}.

From our derivation of $F_k(x)$ in Sec.~\ref{subsec:tmss_coherence}, we know that the sum $\sum_n [n^{\underline{k}}]^2 x^n$ evaluates to $P_k(x)/(1-x)^{2k+1}$. This derivation of the sum is general for any parameter. Thus, for the off-diagonal case, we can make the algebraic substitution $x \to z$:
\begin{align}
\sum_{n=0}^\infty [n^{\underline{k}}]^2 z^n = \frac{P_k(z)}{(1-z)^{2k+1}}.
\end{align}
Multiplying this result by its prefactor $\frac{1}{\cosh r \cosh s}$ gives the final cross-mode kernel. A parallel argument for the single-mode kernel $\langle \zeta | (a^\dagger)^k a^k | \xi \rangle$, which derives from the simpler sum $\sum_n n^{\underline{k}} z^n = k! z^k / (1-z)^{k+1}$, yields its corresponding expression. This demonstrates that the off-diagonal kernels share the same functional form as their diagonal counterparts, evaluated at $z$ instead of $x$. We thus obtain all the essential off-diagonal kernels:
\begin{align}
\langle \zeta | (a^\dagger)^k a^k | \xi \rangle
= \frac{k!}{\cosh r\,\cosh s}\,\frac{z^{\,k}}{(1-z)^{k+1}},
\qquad
\langle \zeta | (a^\dagger b^\dagger)^k (ab)^k | \xi \rangle
= \frac{1}{\cosh r\,\cosh s}\,\frac{P_k(z)}{(1-z)^{2k+1}},
\label{eq:tmjs_kernels_recalled}
\end{align}
where $P_k(u)$ is the squeezing polynomial family defined in Eq.~\eqref{eq:Fk_final_poly}.

Before calculating the $k$-th order coherences, we must first find the mean photon number, which is required for normalization. By applying the bilinear rule \eqref{eq:tmjs_master_bilinear} to the operator $\hat O = a^\dagger a$ (using the $k=1$ case of the kernels), we find the mean photon number in either mode (the state is symmetric under $a \leftrightarrow b$):
\begin{align}
\langle a^\dagger a\rangle_\Psi
=|\chi|^2\,\bar n_r+|\eta|^2\,\bar n_s
+\frac{2|\chi||\eta|}{\cosh r\,\cosh s}\,
\Re\left[e^{-i\delta}\,\frac{z}{(1-z)^2}\right],
\qquad
\langle b^\dagger b\rangle_\Psi=\langle a^\dagger a\rangle_\Psi.
\label{eq:tmjs_mean_for_norm}
\end{align}

We can now assemble the full expression for the single–mode \(k\)-th order factorial moment, $\langle (a^\dagger)^k a^k\rangle_\Psi$. Applying the bilinear rule \eqref{eq:tmjs_master_bilinear} and substituting the diagonal results from Eq. \eqref{eq:gka_tmss} and the off-diagonal kernel from \eqref{eq:tmjs_kernels_recalled} yields:
\begin{align}
\langle (a^\dagger)^k a^k\rangle_\Psi
= k!\,\Big(|\chi|^2\bar n_r^{\,k}+|\eta|^2\bar n_s^{\,k}\Big)
+\frac{2|\chi||\eta|\,k!}{\cosh r\,\cosh s}\,
\Re\left[e^{-i\delta}\,\frac{z^{\,k}}{(1-z)^{k+1}}\right].
\label{eq:tmjs_single_fact_k}
\end{align}
The normalized single-mode coherence $g^{(k)}_a$ is then obtained by dividing this moment by the $k$-th power of the mean photon number:
\begin{align}
g^{(k)}_{a}(\Psi)
=\frac{\langle (a^\dagger)^k a^k\rangle_\Psi}
       {\big(\langle a^\dagger a\rangle_\Psi\big)^{k}},
\quad\text{with } \langle a^\dagger a\rangle_\Psi \text{ from \eqref{eq:tmjs_mean_for_norm}}.
\label{eq:tmjs_gka_def}
\end{align}

Similarly, we compute the cross–mode \(k\)-th order factorial moment, $\langle (a^\dagger b^\dagger)^k(ab)^k\rangle_\Psi$. Using the diagonal results from Eq. \eqref{eq:Fk_final_poly} and the cross-mode kernel from \eqref{eq:tmjs_kernels_recalled}, the bilinear rule gives:
\begin{align}
\langle (a^\dagger b^\dagger)^k(ab)^k\rangle_\Psi
=|\chi|^2\,\frac{P_k(x)}{(1-x)^{2k}}
+|\eta|^2\,\frac{P_k(y)}{(1-y)^{2k}}
+\frac{2|\chi||\eta|}{\cosh r\,\cosh s}\,
\Re\left[e^{-i\delta}\,\frac{P_k(z)}{(1-z)^{2k+1}}\right].
\label{eq:tmjs_cross_fact_k}
\end{align}
The normalized cross–coherence $g^{(k)}_{ab}$ is found by normalizing this moment by the $2k$-th power of the mean photon number (since $\langle a^\dagger a \rangle = \langle b^\dagger b \rangle$):
\begin{align}
g^{(k)}_{ab}(\Psi)
=\frac{\langle (a^\dagger b^\dagger)^k (ab)^k\rangle_\Psi}
       {\big(\langle a^\dagger a\rangle_\Psi\big)^{2k}},
\quad\text{with } \langle a^\dagger a\rangle_\Psi \text{ from \eqref{eq:tmjs_mean_for_norm}}.
\label{eq:tmjs_gkab_def}
\end{align}

A key feature of the TMJS is the ability to control these coherences via the Janus phase \(\delta\). Every interference term in the expressions above shares a universal mathematical form:
\begin{align}
2|\chi||\eta|\,\Re\big[e^{-i\delta} f_k(z)\big]
=2|\chi||\eta|\,|f_k(z)|\,\cos\big(\varphi_k-\delta\big),
\end{align}
where $f_k(z)$ is the complex-valued kernel for the observable of interest. Specifically, \(f_k(z)=z^{\,k}(1-z)^{-(k+1)}\) for the single-mode coherence \eqref{eq:tmjs_single_fact_k}, and
\(f_k(z)=P_k(z)(1-z)^{-(2k+1)}\) for the cross-mode coherence \eqref{eq:tmjs_cross_fact_k}. The phase of this kernel is denoted by \(f_k(z)=|f_k(z)|e^{i\varphi_k}\).
From this, we see that constructive or destructive interference can be precisely selected by tuning the external phase \(\delta\):
\begin{align}
\delta=\varphi_k \ \ (\mathrm{mod}\,2\pi)
\qquad
\big(\text{respectively } \delta=\varphi_k+\pi\big).
\end{align}
This provides a direct experimental knob for "steering" the quantum statistics of the state.

\section{Visualization of phase steering, coherence landscapes, and phase–space structure} \label{sec:vis}

Having derived the full coherence functions for the TMJS in the preceding subsections, we now visualize and discuss their behavior. The key feature of the TMJS is the ability to "steer" the $k$-th order coherence via the Janus phase $\delta$ and the relative squeezing phase $\Delta = \theta - \phi$. We will assume a symmetric squeezing scenario ($r=s$) to isolate the effects of interference.

\begin{figure}[t]
\centering
\includegraphics[width=.9\textwidth]{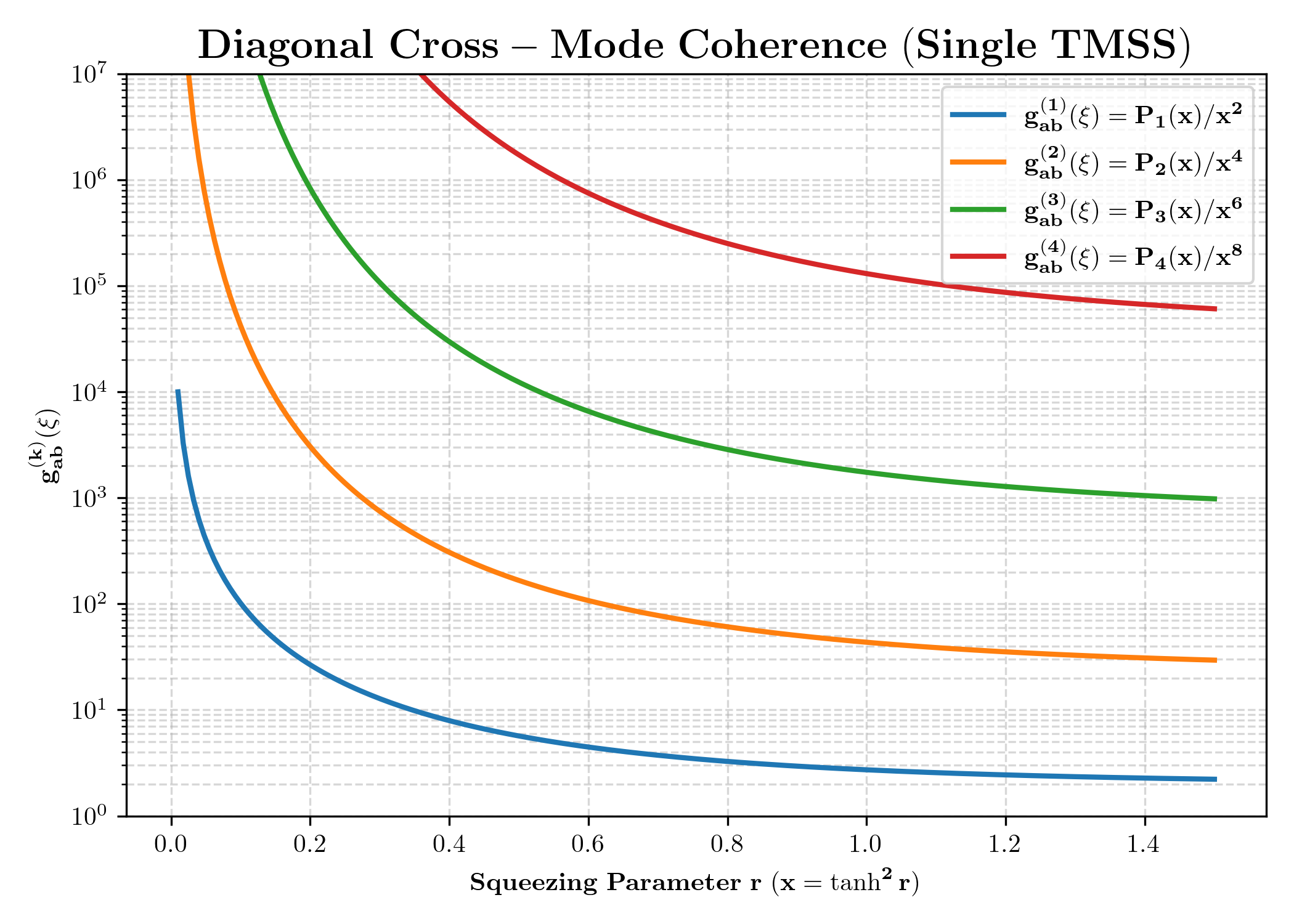}
\caption{\textbf{Baseline (no interference):} Diagonal cross–mode coherences for a \emph{single} TMSS, $g_{ab}^{(k)}(\xi)=P_k(x)/x^{2k}$, where $P_k$ are the squeezing polynomials. The four curves (from bottom to top) show $k=1,2,3,4$ on a log scale as a function of $r$ ($x=\tanh^2 r$). As $r\to0$ one has $P_k(x)\sim (k!)^2 x^k$ so $g_{ab}^{(k)}\sim (k!)^2 x^{-k}\propto r^{-2k}$, explaining the steep small–$r$ divergence that sets the reference scale for all Janus effects.}
\label{fig:fig_pk_polynomials}
\end{figure}

\begin{figure}[t]
\centering
\includegraphics[width=.98\textwidth]{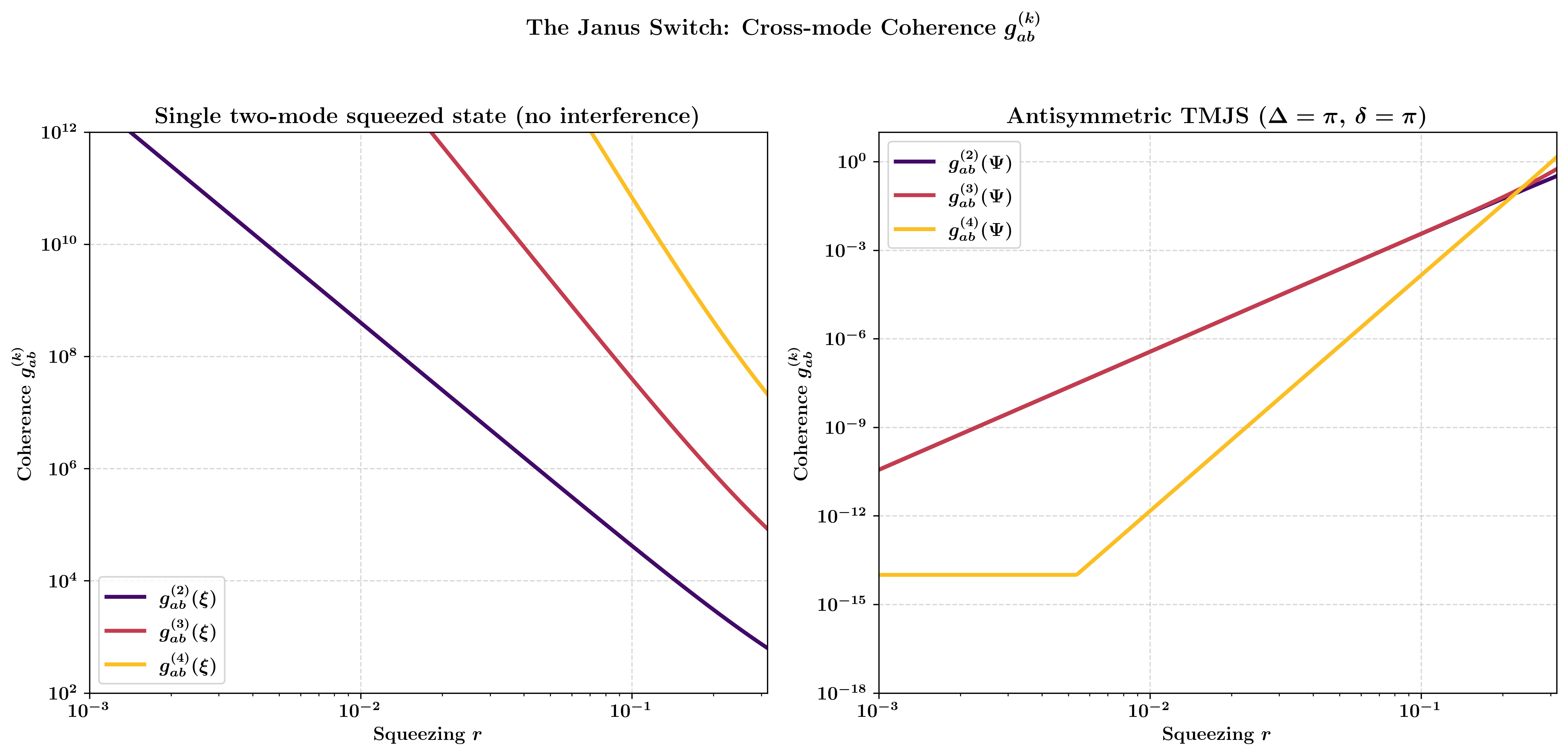}
\caption{\textbf{Interference turns the knob:} \emph{Left}—the same single–TMSS divergences from Fig.~\ref{fig:fig_pk_polynomials} on a log–log axis. \emph{Right}—the \emph{antisymmetric} TMJS ($\Delta=\pi,\ \delta=\pi$) where destructive interference suppresses cross–mode coherences by many orders of magnitude at small $r$. The opposite slopes of the two panels emphasize that the Janus superposition replaces the $r^{-2k}$ blow–up by deep cancellations that can drive $g_{ab}^{(k)}(\Psi)\ll1$.}
\label{fig:janus_coherence_switch}
\end{figure}

\begin{figure}[t]
\centering
\includegraphics[width=.98\textwidth]{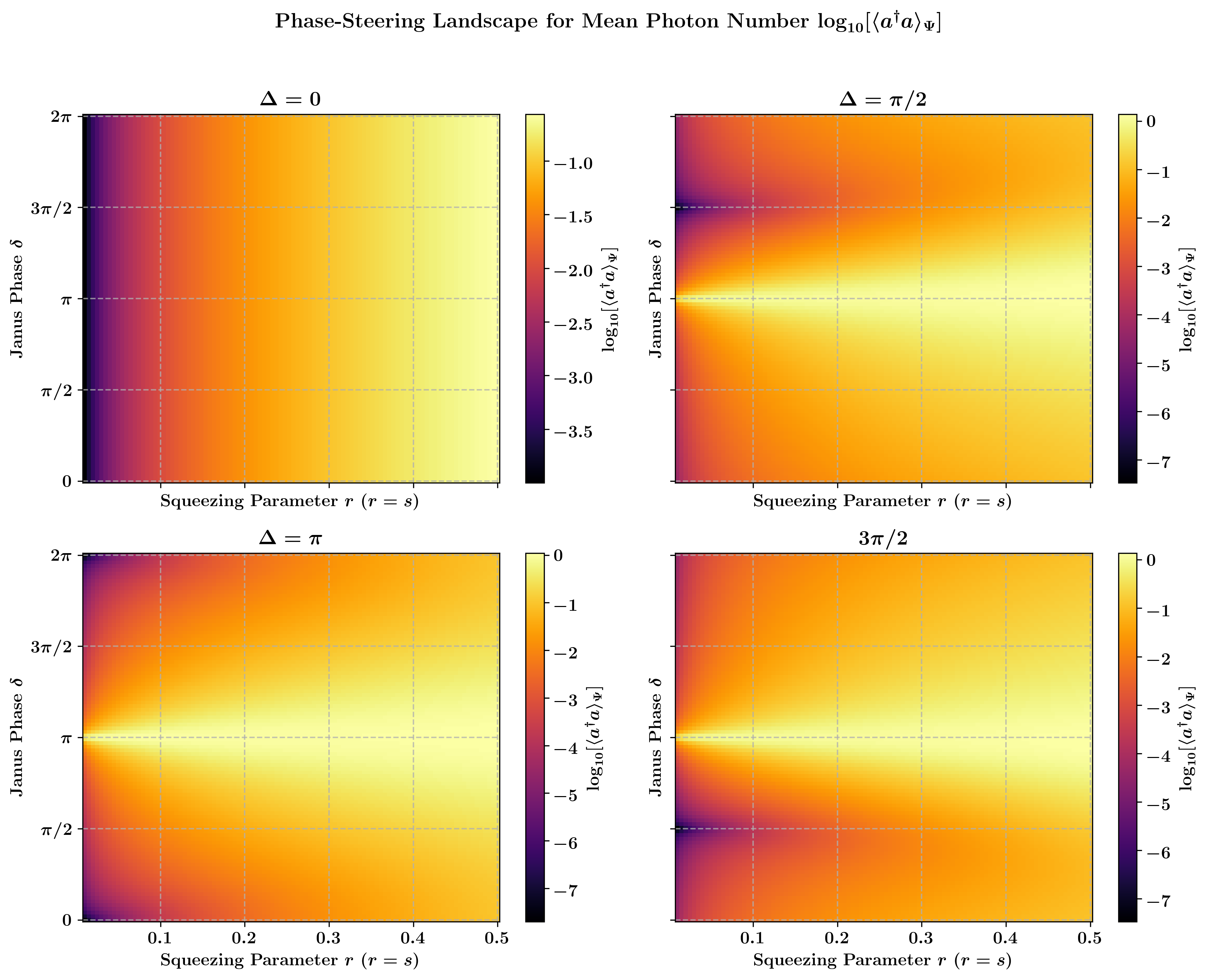}
\caption{\textbf{Mean brightness landscape} $\log_{10}\!\big(\langle a^\dagger a\rangle_\Psi\big)$ over $(r,\delta)$ for $\Delta\in\{0,\pi/2,\pi,3\pi/2\}$ with $r=s$. This quantity is the \emph{denominator seed} that ultimately enters all normalized single–mode coherences. The dark trench at $\delta=\pi$ for $\Delta=\pi$ pinpoints maximal destructive interference in the marginal, while the bright lobes for $\Delta=\pi/2$ reflect quadrature–skewed constructive terms. Reading this figure first helps interpret where single–mode $g_a^{(k)}(\Psi)$ will be most strongly enhanced or suppressed.}
\label{fig:brightness_landscape}
\end{figure}

\begin{figure}[t]
\centering
\includegraphics[width=.98\textwidth]{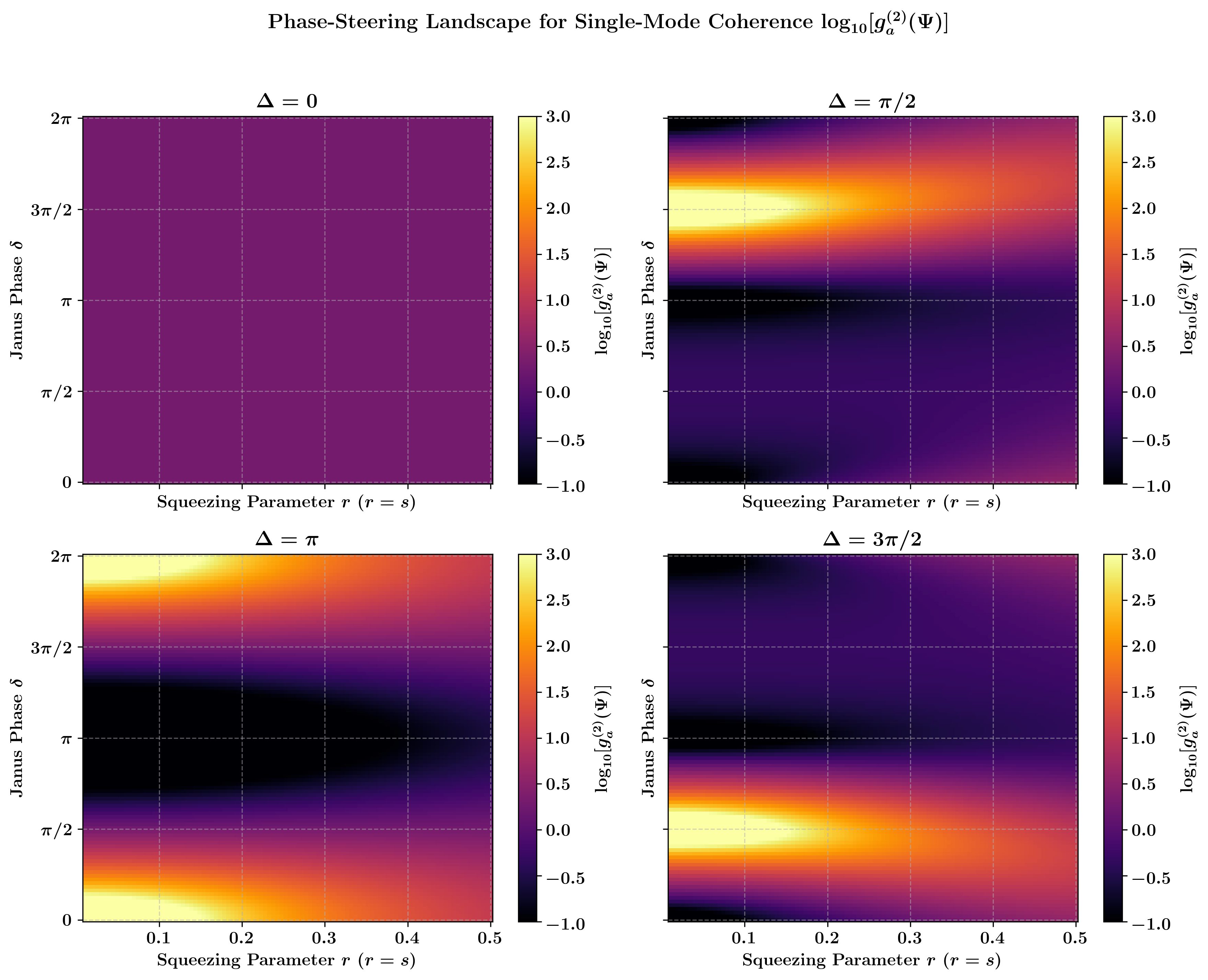}
\caption{\textbf{Single–mode coherence $g_a^{(2)}(\Psi)$:} $\log_{10}$ heat maps vs.\ $(r,\delta)$ at fixed $\Delta$. When $\Delta=0$ the landscape is essentially flat: the two constituents are in phase and the normalized second–order coherence reduces to its single–state value with negligible $\delta$–dependence. For $\Delta=\pi$ a deep valley opens along $\delta=\pi$, mirroring the dark trench in Fig.~\ref{fig:brightness_landscape}; moving away from $\delta=\pi$ restores super–Poissonian structure. Quarter–period shifts ($\Delta=\pi/2,3\pi/2$) rotate the bright–dark lobes as expected from the $z=e^{i\Delta}\tanh r\tanh s$ kernel.}
\label{fig:single_mode_g2_landscape}
\end{figure}

\begin{figure}[t]
\centering
\includegraphics[width=.98\textwidth]{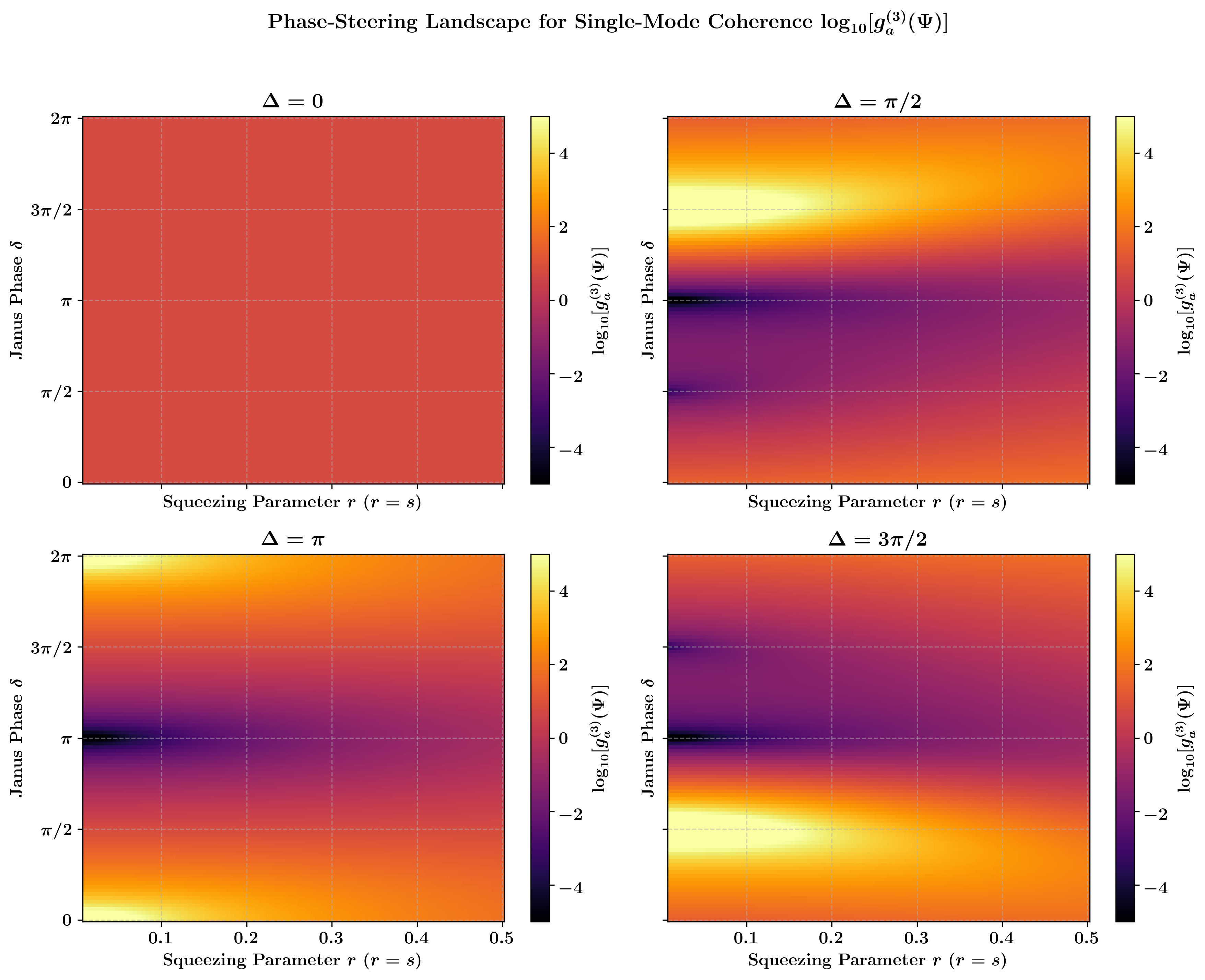}
\caption{\textbf{Single–mode coherence $g_a^{(3)}(\Psi)$:} Compared to $k=2$, phase selectivity is sharper: the nodal trench at $(\Delta,\delta)=(\pi,\pi)$ is narrower and the surrounding ridges are higher. This is consistent with the higher–order cross kernels $P_k(z)/(1-z)^{2k+1}$: larger $k$ amplifies both interference gains and cancellations, so the dynamic range of $\log_{10}g_a^{(3)}$ increases while its extrema track those in Fig.~\ref{fig:brightness_landscape}.}
\label{fig:single_mode_g3_landscape}
\end{figure}

\begin{figure}[t]
\centering
\includegraphics[width=.98\textwidth]{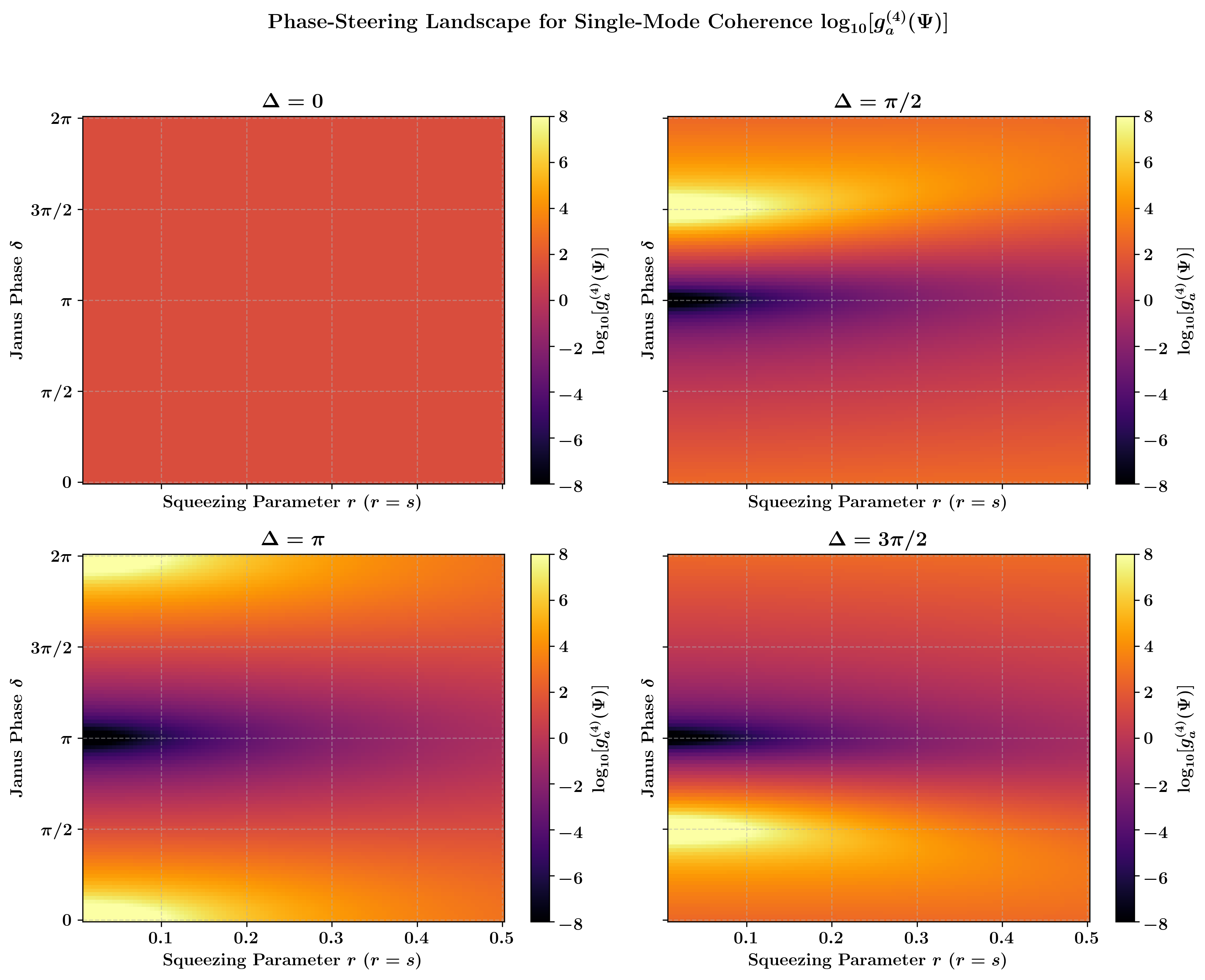}
\caption{\textbf{Single–mode coherence $g_a^{(4)}(\Psi)$:} The $k=4$ landscape accentuates the same geometry: flat for $\Delta=0$, quadrature–rotated lobes for $\Delta=\pi/2,3\pi/2$, and a pronounced cancellation trench for $\Delta=\pi$ centered at $\delta=\pi$. Increasing $k$ therefore offers a powerful phase–contrast knob: higher moments are more sensitive to phase misalignment but also deliver deeper minima near optimal destructive settings.}
\label{fig:single_mode_g4_landscape}
\end{figure}

\begin{figure}[t]
\centering
\includegraphics[width=.98\textwidth]{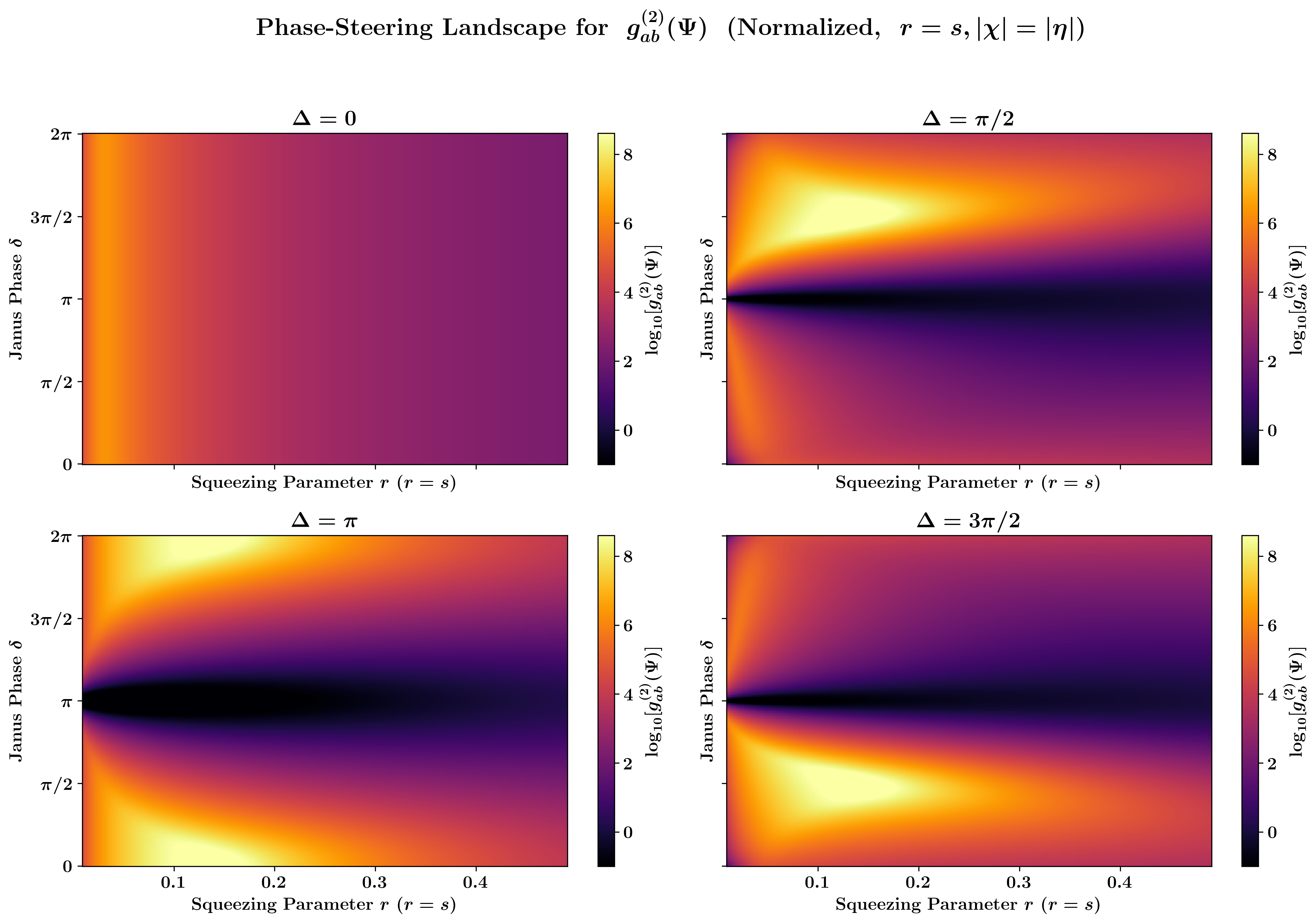}
\caption{\textbf{Cross–mode coherence $g_{ab}^{(2)}(\Psi)$ (normalized, $r=s$, $|\chi|=|\eta|$):} A full $(r,\delta)$ sweep for the four $\Delta$ values. In contrast to single–mode moments, $g_{ab}^{(2)}$ directly exposes the phase of the \emph{pair} kernel. The global minimum rides the antisymmetric setting $(\Delta,\delta)=(\pi,\pi)$; the brightest ridge aligns with constructive $e^{-i\delta}\langle \zeta|\xi\rangle$ phases. These maps, together with Fig.~\ref{fig:janus_coherence_switch}, quantify the “Janus switch’’ between correlation blow–up (single TMSS) and correlation quenching (TMJS).}
\label{fig:fig_r_delta_landscape_k2}
\end{figure}

\begin{figure}[t]
\centering
\includegraphics[width=.98\textwidth]{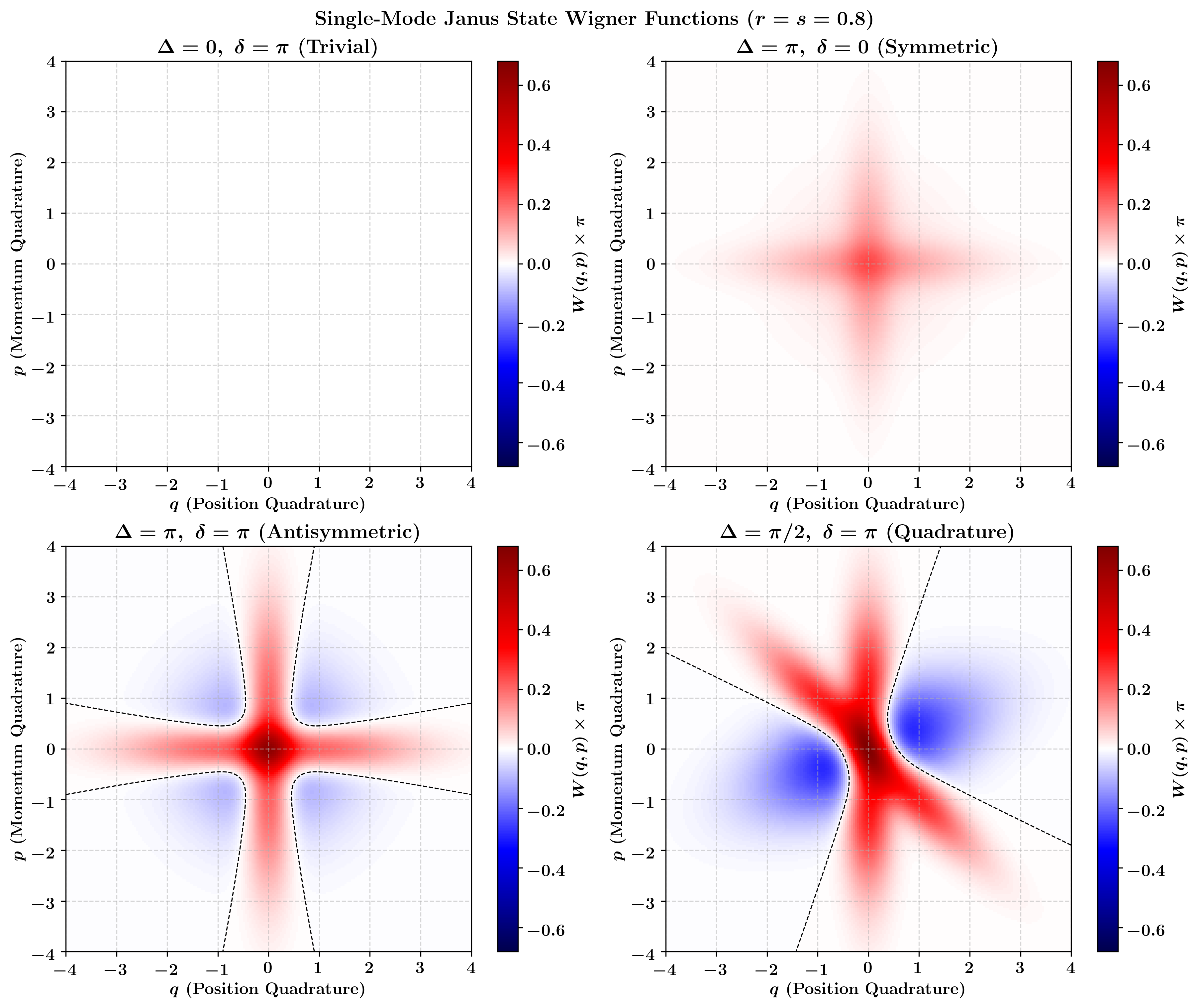}
\caption{\textbf{Phase–space view (single–mode Janus Wigner functions, $r=s=0.8$):} Four panels illustrate how interference reshapes the quasi–probability distribution. For $\Delta=0,\ \delta=\pi$ the gallery is nearly featureless (trivial superposition); for $\Delta=\pi,\ \delta=0$ (symmetric) one observes four–lobe structures from constructive overlap in complementary quadratures; the \emph{antisymmetric} case $\Delta=\pi,\ \delta=\pi$ exhibits pronounced negative regions bounded by the dashed $W=0$ contours, reflecting the deep cancellations concurrently seen in Figs.~\ref{fig:single_mode_g3_landscape}--\ref{fig:fig_r_delta_landscape_k2}. The quadrature case $\Delta=\pi/2,\ \delta=\pi$ rotates these interference fringes as predicted by the phase of $z$.}
\label{fig:janus_wigner_gallery}
\end{figure}

We visualize how the two control phases---the relative squeezing phase $\Delta$ and the Janus phase $\delta$---steer single–mode and cross–mode coherences, and how these correlates appear directly in the phase–space (Wigner) picture. Throughout, $r$ and $s$ denote the squeezing magnitudes of the constituent two–mode squeezed vacua, $x=\tanh^2 r$, $y=\tanh^2 s$, and $z=\tanh r\,\tanh s\,e^{i\Delta}$. Unless otherwise stated we work in the symmetric case $r=s$ and equal weights $|\chi|=|\eta|$, with proper normalization enforced by the TMSS overlap.

We now analyze the figures in detail to build a complete picture of the TMJS, from its baseline components to its full, phase-steered behavior.

Figure \ref{fig:fig_pk_polynomials} establishes the baseline behavior for the system's components. It plots the diagonal cross–mode coherence, $g_{ab}^{(k)}(\xi)$, for a \emph{single} TMSS, corresponding to the "which-path available" or non-interfering limit. As shown for $k=1, 2, 3,$ and $4$, all orders of coherence strongly diverge as the squeezing parameter $r$ approaches zero. This $g_{ab}^{(k)} \sim r^{-2k}$ divergence, explained in the caption, is characteristic of the strong multi-photon bunching inherent to a single TMSS. This divergent behavior is the reference against which the Janus interference effects will be compared.

Figure \ref{fig:janus_coherence_switch} provides the central thesis of this work. It directly contrasts the baseline behavior of a single TMSS (left panel) with the fully interfering, antisymmetric TMJS (right panel, set to the optimal suppression phases $\Delta=\pi, \delta=\pi$). The left panel recasts the divergence from Fig. \ref{fig:fig_pk_polynomials} on a log-log scale. The right panel, in stark contrast, shows the result of "which-path erasure." The quantum interference is so effective that the statistical behavior is completely inverted: all higher-order cross-mode coherences, $g_{ab}^{(k)}(\Psi)$, are driven to zero by many orders of magnitude. The opposite slopes of the lines in the two panels provide a clear visualization of the "Janus switch," which tunes the system from infinite bunching to deep suppression.

Figure \ref{fig:brightness_landscape} is a critical control plot that shows the mean photon number of the single-mode marginal, $\langle a^\dagger a\rangle_\Psi$. This quantity forms the denominator for all normalized single-mode statistics ($g_a^{(k)}$). The plot reveals that the brightness itself is highly steerable by the phases. Crucially, in the primary suppression region (e.g., $\Delta=\pi, \delta=\pi$), the mean photon number is clearly non-zero and, in fact, is *maximized*. This is a critical finding: the suppression of correlations is a true quantum interference effect, not a trivial consequence of the state having no photons.

The set of figures \ref{fig:single_mode_g2_landscape}, \ref{fig:single_mode_g3_landscape}, and \ref{fig:single_mode_g4_landscape} illustrates the universal nature of the Janus switch for the single-mode marginal statistics $g_a^{(k)}(\Psi)$.
\begin{itemize}
    \item The $\Delta=0$ panel in all three figures is a flat, uniform color. The values correspond to $\log_{10}(2!) \approx 0.3$, $\log_{10}(3!) \approx 0.78$, and $\log_{10}(4!) \approx 1.38$. This perfectly confirms our derivation that the single-mode marginal of a single, non-interfering TMSS is perfectly thermal, with $g_a^{(k)} = k!$.
    \item The $\Delta \neq 0$ panels demonstrate that as soon as a non-trivial relative squeezing phase is introduced, the Janus phase $\delta$ becomes an active control knob. For all $k$, a deep valley of suppression (dark purple/black) opens up, centered at $\delta=\pi$. Conversely, regions of extreme enhancement (bright yellow/orange) appear at $\delta=0$ and $2\pi$.
    \item Comparing the three figures shows that this pattern is universal. The suppression valleys become progressively deeper and the enhancement hills higher as $k$ increases, demonstrating that higher-order coherences are even more sensitive to the phase-steering.
\end{itemize}

Figure \ref{fig:fig_r_delta_landscape_k2} returns to the cross-mode coherence, $g_{ab}^{(2)}(\Psi)$, and maps its full landscape. The structure mirrors that of the single-mode plots, confirming that the Janus phases steer both local ($a^\dagger a$) and non-local pair ($a^\dagger b^\dagger$) correlations. The $\Delta=0$ panel is flat (diverging as $r \to 0$), while all other panels show a deep suppression valley at $\delta=\pi$. This figure is the full 2D map from which the 1D slice in Fig. \ref{fig:janus_coherence_switch} (right panel) is taken, quantifying the full range of the "Janus switch" for two-photon correlations.

Finally, Figure \ref{fig:janus_wigner_gallery} reveals the underlying physical mechanism for this behavior by plotting the Wigner function of the simpler \emph{single-mode} Janus analogue. The four panels show the four key phase configurations. The symmetric case (b, $\Delta=\pi, \delta=0$) shows constructive interference and is positive everywhere. In contrast, the antisymmetric case (c, $\Delta=\pi, \delta=\pi$)—which corresponds to the optimal suppression setting in all previous plots—exhibits dramatic destructive interference, creating a large, unmistakable region of Wigner negativity (blue, bounded by dashed lines) at the origin. This negativity is the definitive phase-space signature of the state's non-Gaussianity and is the physical mechanism responsible for the suppression of $g^{(k)}$ correlations.

To \emph{quench} correlations, choose the antisymmetric setting $(\Delta,\delta)=(\pi,\pi)$ and moderate $r$ so that the TMSS overlap is large enough to interfere but not so large that residual diagonal terms dominate; to \emph{boost} correlations, steer towards $\delta\approx0$ with $\Delta$ tuned to place the bright lobe of Fig.~\ref{fig:brightness_landscape} at the target $r$. Higher–order moments ($k=3,4$) magnify the contrast, at the price of narrower phase tolerances.

\section{Physical Realization and QFT Analogues}
\label{sec:physical_realization}
This section clarifies what the Unruh/Rindler correspondence does---and does not---buy us for preparing a TMJS. The key point is that an observer's acceleration, by itself, does not create different \emph{field states}; it only changes the observer's notion of particles and their physical response to the existing vacuum. By contrast, \emph{time-dependent dynamics of the field} (e.g., accelerated boundaries, parametric driving, or cosmological pair creation) are true state-generating processes. These dynamics \emph{do} produce the distinct TMSS resources that can be coherently superposed to realize a TMJS.

\subsection{What Unruh/Rindler does and does not do}
\label{subsec:unruh_clarified}

Let $\eta$ denote the boost time in a Rindler wedge and $-i\partial_\eta$ the corresponding generator. The Minkowski vacuum admits the exact continuous-mode TMSS factorization across the L/R wedges,
\begin{align}
\ket{0_M}
= \frac{1}{\sqrt{\mathcal Z}}
\exp\!\Bigg[\int_{0}^{\infty}\! d\omega\; e^{-\pi\omega}\, b_{L\omega}^\dagger b_{R\omega}^\dagger \Bigg]
\ket{0_L}\otimes\ket{0_R},
\label{eq:Minkowski_as_TMSS}
\end{align}
where $\omega>0$ is the \emph{dimensionless} boost frequency (the eigenvalue of $-i\partial_\eta$). No ``acceleration parameter'' appears in \eqref{eq:Minkowski_as_TMSS}; the state is fixed by Poincar\'e invariance. To use the discrete-mode formulas of Secs.~\ref{subsec:tmss_coherence}--\ref{subsec:tmjs_coherence}, one wavepacketizes the boost-frequency continuum into narrow, non-overlapping packets; within each packet the weight $e^{-\pi\omega}$ is approximately constant, yielding a per-packet TMSS.

Uniformly accelerated worldlines have proper time $\tau$ related to $\eta$ by $\tau=\eta/a_0$ (with $a_0$ the proper acceleration). Detectors coupled along such worldlines physically observe a KMS thermal bath at temperature $T=a_0/2\pi$. However, this temperature characterizes the detector's physical \emph{response} to the static $\ket{0_M}$ state; it is not a signal that a new global field state has been prepared.

This distinction is fundamental. As rigorously shown in \cite{Azizi2023JHEP}, the Rindler vacuum is unique and independent of the observer's specific acceleration. The Bogoliubov transformation between Rindler operators defined for two different accelerations, $a$ and $a'$, is ``trivial''---it does not mix creation and annihilation operators. This proves that all Rindler observers, regardless of their acceleration scale, are probing the \emph{same} vacuum.

\begin{remark}
The Rindler  vacuum is unique. The Minkowski vacuum's L/R TMSS decomposition \eqref{eq:Minkowski_as_TMSS} is a fixed, intrinsic property of the state. Superposing different observer accelerations cannot yield a superposition of different \emph{field} TMSSs (a TMJS), precisely because all such observers share the identical vacuum.
\end{remark}

Consequently, any physical construction of a TMJS must be based on \emph{dynamics that change the field state itself}, not merely on a change of observer.

\subsection{Operational routes that \emph{do} prepare distinct TMSSs}
\label{subsec:operational_routes}

Having established that a physical process---not merely an observer---is required to generate new field states, we now identify a family of dynamics that provides the necessary ``bona fide'' TMSS resources. These are true, multi-mode squeezed states created from the vacuum by a physical interaction. The central proposal is that by coherently superposing two distinct \emph{histories} of such a process, one can generate a TMJS in the sense of Sec.~\ref{subsec:tmjs_coherence}.

\subsubsection*{(i) Accelerated boundaries / dynamical Casimir effect}

The most general and physically direct mechanism is the Dynamical Casimir Effect (DCE). In this scenario, a time-dependent, perfectly reflecting boundary (i.e., an accelerating mirror) with a worldline $v=p(u)$ actively injects energy into the field. This physical disturbance induces a non-trivial Bogoliubov transformation that \emph{creates} correlated pairs of real quanta from the initial in-vacuum.

This Bogoliubov map is, in general, a complex, non-diagonal transformation that mixes all field modes. However, the Bloch--Messiah reduction (or Schmidt decomposition) theorem provides a powerful simplification. It guarantees that there exists a new, customized "Schmidt packet" basis (labeled by a discrete index $k$) in which this complex multimode out-state factorizes into a simple tensor product of independent Two-Mode Squeezed States (TMSSs).
Thus, a single, complete history of mirror motion, $\mathcal T_1$, generates a specific family of TMSSs, one for each Schmidt mode $k$: $\{\ket{\xi_k^{(1)}} = S_2(\xi_k^{(1)})\ket{0}_k\}$.

Now, consider two \emph{distinct} drive histories, $\mathcal T_1$ and $\mathcal T_2$. These could represent different acceleration profiles, different drive durations, or (as discussed in Sec.~\ref{subsec:dce}) more complex "Ramsey-like" squeeze--dwell--unsqueeze sequences. Each history, when run independently, generates its own unique family of TMSSs: $\{\xi_k^{(1)}\}$ and $\{\xi_k^{(2)}\}$, respectively.

To create the TMJS, we coherently superpose these histories. We can imagine coupling the boundary's motion to an ancilla qubit. We first prepare the ancilla in a superposition $\chi\ket{1} + \eta e^{i\delta}\ket{2}$. Then, a controlled-unitary operation enacts history $\mathcal T_1$ if the ancilla is in state $\ket{1}$ and history $\mathcal T_2$ if it is in state $\ket{2}$. By post-selecting the ancilla in a superposition state (e.g., $\ket{+} \propto \ket{1} + \ket{2}$), we "erase" the which-history information. This projection leaves the quantum field itself in the desired coherent superposition. Assuming an idealized case where the Schmidt bases of both histories are perfectly mode-matched, the final field state factorizes, preparing, mode by mode:
\begin{align}
\ket{\Psi_k} \propto \chi\,\ket{\xi_k^{(1)}} + \eta\,e^{i\delta}\,\ket{\xi_k^{(2)}},
\end{align}
which is precisely the definition of a TMJS for each Schmidt mode $k$. The physically generated parameters $r_k^{(j)}$ and $\vartheta_k^{(j)}$ from the two trajectories map directly to the abstract parameters of our master formulas. By identifying $x_k=\tanh^2 r_k^{(1)}$, $y_k=\tanh^2 r_k^{(2)}$, and the crucial interference term
\begin{align}
z_k = e^{i(\vartheta_k^{(1)}-\vartheta_k^{(2)})}\,\tanh r_k^{(1)}\,\tanh r_k^{(2)},
\qquad |z_k|<1,
\end{align}
all the coherence calculations of Secs.~\ref{subsec:tmss_coherence} and \ref{subsec:tmjs_coherence} become directly applicable to this physical, per-mode construction.

\subsection{Physical Realization via Superposed Boundaries}
\label{subsec:dce}
The abstract formalism of the TMJS finds a direct and physically-grounded realization in the Dynamical Casimir Effect (DCE). The DCE describes the creation of real particles from the vacuum due to the motion of a boundary (e.g., an accelerating mirror). This section builds this model, starting with a crucial distinction between the observer-dependent Unruh effect and the state-generating DCE.

\subsubsection{Unruh Effect vs. Dynamical Casimir Effect}
\label{subsubsec:unruh_vs_dce}
It is crucial to separate the \emph{quantum state of the field} from an \emph{observer’s definition of particles}. The two canonical effects highlight this difference:

\begin{enumerate}
    \item \textbf{Unruh Effect (Accelerated Observer, Static Vacuum):} The global field is prepared in the Minkowski vacuum, $\ket{0_M}$. An accelerated observer, by virtue of their motion, naturally uses Rindler modes to describe the field. In this Rindler basis, the \emph{same} Minkowski vacuum state $\ket{0_M}$ is mathematically equivalent to a two-mode squeezed state (TMSS) entangling the right and left Rindler wedges.
    
    This observer will physically detect a thermal bath of particles. However, this describes the \emph{response} of a localized, accelerated probe, which draws energy from the agent causing its acceleration to click. It is not a \emph{state-generating process} that alters the global quantum field. An inertial observer in the same spacetime still sees the empty, unchanged Minkowski vacuum. As shown in \cite{Azizi2023JHEP}, the Rindler vacuum is unique and independent of the acceleration scale, confirming that changing the observer's acceleration does not create a new field state.

    \item \textbf{Dynamical Casimir Effect (Inertial Observer, Dynamic Boundary):} Here, the observer remains inertial, but the boundary (the mirror) is accelerated. This time-dependent boundary injects real energy \emph{into the field itself}. This motion induces a nontrivial Bogoliubov transformation on the \emph{in}-modes, one with a non-zero $\beta$ coefficient that mixes creation and annihilation operators. The resulting \emph{out}-state is no longer the Minkowski vacuum. It is a new quantum state, one populated with real, correlated quanta that are detectable by \emph{any} observer, including inertial ones. This is a true state-generating, physical process.
\end{enumerate}

With this distinction, we can properly define the "vacuum" for the accelerating mirror scenario. For a perfect reflector with a worldline $v=p(u)$ in $1{+}1$ dimensions, the starting point is the \emph{in-vacuum}: the state on the past null infinity, $\mathscr{I}^-$, which contains no incoming quanta. If the mirror was static in the far past, this is just $\ket{0_M}$. The subsequent acceleration of the mirror implements a dynamical $SU(1,1)$ Bogoliubov transformation that \emph{creates} correlated pairs, driving this well-defined in-vacuum to a nontrivial out-state populated with particles.

In the massless case, this physical creation of energy is explicit. The non-inertial boundary motion acts as a direct source of radiation, generating a detectable energy flux given by the Schwarzian derivative of the trajectory:
\begin{align}
\langle T_{uu}\rangle_{\rm out}=-\frac{1}{24\pi}\,\{p(u),u\}.
\end{align}

This scattering process is a general Bogoliubov transformation. For a non-stationary trajectory, the map is not diagonal in the standard monochromatic frequency basis. Instead, it mixes \emph{all} incoming frequencies ($\Omega'$) to produce a single outgoing frequency ($\Omega$). Let $a_\Omega^{\rm in}, b_\Omega^{\rm in}$ be operators for incoming modes. The outgoing operators are given by complex integral transforms:
\begin{align}
a_\Omega^{\rm out}
=\int_{0}^{\infty}\! d\Omega'\,
\Big[\alpha_{\Omega\Omega'}\,a_{\Omega'}^{\rm in}+\beta_{\Omega\Omega'}\,b_{\Omega'}^{\rm in\,\dagger}\Big],
\qquad
b_\Omega^{\rm out}
=\int_{0}^{\infty}\! d\Omega'\,
\Big[\tilde\alpha_{\Omega\Omega'}\,b_{\Omega'}^{\rm in}+\tilde\beta_{\Omega\Omega'}\,a_{\Omega'}^{\rm in\,\dagger}\Big],
\label{eq:mirror_bog_kernel}
\end{align}
which must satisfy constraints to preserve the canonical commutation relations. This general, "messy" linear transformation can always be diagonalized. By the Bloch--Messiah (or Schmidt-mode) decomposition, there always exists a new, discrete basis of orthonormal wave-packets---the "Schmidt modes" $k$---that are custom-built for this specific trajectory.

In this special Schmidt basis, denoted $(\tilde a_k^{\rm in}, \tilde b_k^{\rm in})$ and $(\tilde a_k^{\rm out}, \tilde b_k^{\rm out})$, the complex integral transformation \eqref{eq:mirror_bog_kernel} simplifies dramatically. It reduces to a set of independent, $2\times 2$ $SU(1,1)$ maps, one for each Schmidt mode $k$, with no mixing between different $k$:
\begin{align}
\tilde a_k^{\rm out}=\cosh r_k\,\tilde a_k^{\rm in}+e^{i\vartheta_k}\sinh r_k\,\tilde b_k^{\rm in\,\dagger},
\qquad
\tilde b_k^{\rm out}=e^{-i\vartheta_k}\sinh r_k\,\tilde a_k^{\rm in\,\dagger}+\cosh r_k\,\tilde b_k^{\rm in}.
\label{eq:mirror_schmidt_map}
\end{align}
This provides the crucial link: the in-vacuum $|0\rangle_{\rm in} = \bigotimes_{k} |0\rangle_{{\rm in}, k}$ evolves to a tensor product of independent two-mode squeezed states, one for each Schmidt mode $k$:
\begin{align}
|\Psi\rangle_{\rm out} = \bigotimes_{k} |\xi_k\rangle = \bigotimes_{k} S_2(\xi_k)|0\rangle_{{\rm in}, k},\qquad
\xi_k=r_k e^{i\vartheta_k}.
\end{align}
This establishes the rigorous connection: a single, generic accelerating mirror trajectory produces a set of discrete, independent TMSSs, not in the frequency basis, but in its own unique Schmidt-mode wave-packet basis.

This direct link provides a clear path to generating a TMJS. If a \emph{single} history produces a TMSS, a \emph{superposition} of histories can produce a superposition of TMSSs. Consider two distinct mirror trajectories, $\mathcal{T}_1$ and $\mathcal{T}_2$, which generate two different sets of squeezing parameters, $\{\xi_k^{(1)}\}$ and $\{\xi_k^{(2)}\}$. We can use an ancilla system (e.g., a qubit $\mathsf{G}$) to quantum-mechanically control which history is enacted.

We start by preparing the ancilla in a superposition $\chi|1\rangle_{\mathsf{G}}+\eta\,e^{i\delta}|2\rangle_{\mathsf{G}}$. We then apply a controlled evolution $U_{\rm ctrl}$ that couples the ancilla to the field:
\begin{align}
U_{\rm ctrl}=|1\rangle\!\langle 1|_{\mathsf{G}}\otimes U_1+|2\rangle\!\langle 2|_{\mathsf{G}}\otimes U_2,
\end{align}
where $U_j$ is the unitary evolution corresponding to trajectory $\mathcal{T}_j$. If the ancilla is $\ket{1}$, the mirror performs $\mathcal{T}_1$; if it is $\ket{2}$, it performs $\mathcal{T}_2$. This entangles the ancilla with the field, producing the state $\chi(U_1|0\rangle)\ket{1} + \eta e^{i\delta}(U_2|0\rangle)\ket{2}$.

Finally, we "erase" the which-history information by projecting the ancilla onto a superposition state, such as \(|+\rangle_{\mathsf{G}}\propto |1\rangle_{\mathsf{G}}+|2\rangle_{\mathsf{G}}\). This post-selection disentangles the ancilla and projects the field into the desired coherent superposition:
\begin{align}
|\Psi\rangle_{\rm out}\ \propto\
\chi\,U_1|0\rangle+\eta\,e^{i\delta}\,U_2|0\rangle.
\label{eq:tmjs_mirror_generic}
\end{align}
For this to result in a simple, per-mode TMJS, we must address the issue of the Schmidt basis. Each history $j$ has its own Bloch--Messiah decomposition, $U_j=W^{(j)}\Big[\prod_k S_2(\xi^{(j)}_k)\Big]V^{(j)}$, with passive (beam-splitter-like) mode-mixing unitaries $V^{(j)}, W^{(j)}$ that define its unique Schmidt basis. If the bases for $\mathcal{T}_1$ and $\mathcal{T}_2$ are different (i.e., $V^{(1)}\!\neq\!V^{(2)}$), the resulting state \eqref{eq:tmjs_mirror_generic} is a complex superposition that is not a simple product.

If, however, we assume an idealized case where mode-matching is enforced (i.e., the trajectories are chosen such that $V^{(1)}\!=\!V^{(2)}$ and $W^{(1)}\!=\!W^{(2)}$), the state factorizes into a simple product. This process prepares, \emph{for each Schmidt mode $k$}, a TMJS:
\begin{align}
|\Psi\rangle_{\rm out}\ \propto\ \bigotimes_{k}\Big(\chi\,|\xi_k^{(1)}\rangle+\eta\,e^{i\delta}\,|\xi_k^{(2)}\rangle\Big).
\label{eq:tmjs_mirror_construct}
\end{align}
Assuming this idealized mode-matched case, the parameters of this physical TMJS map directly to the abstract formalism of Sec.~\ref{sec:derivations}. For each mode $k$, we identify the squeezing parameters $\xi_k^{(j)}=r_k^{(j)}e^{i\vartheta_k^{(j)}}$ from each history's Schmidt decomposition and set:
\begin{align}
x_k=\tanh^2 r_k^{(1)},\qquad
y_k=\tanh^2 r_k^{(2)},\qquad
z_k=e^{i(\vartheta_k^{(1)}-\vartheta_k^{(2)})}\tanh r_k^{(1)}\tanh r_k^{(2)}.
\label{eq:xyz_map_mirror}
\end{align}
With this mapping, all our derived formulas for $g^{(k)}$ apply on a mode-by-mode basis. The interference kernel for any observable, $f_k(z_k)=|f_k(z_k)|e^{i\varphi_k}$, can be controlled by the external Janus phase $\delta$ via the term $\cos(\varphi_k-\delta)$, enabling per-mode steering of the higher-order statistics.

A practical question is how to generate a non-trivial phase difference $\vartheta_k^{(1)}-\vartheta_k^{(2)}$, which is required for a complex $z_k$ and thus for non-trivial interference. Single-step squeezing processes often yield real Bogoliubov coefficients $\alpha_k, \beta_k$, resulting in a real $z_k$. To \emph{dial} this phase, we can employ a "Rindler/Mirror Ramsey" sequence by interleaving squeezing with free evolution. In the Bogoliubov two-component picture, a sudden squeeze $Q(r)$, followed by a "dwell time" of free evolution $R(\phi)$, and an "unsqueeze" $Q(-r)$ yields a total transformation:
\begin{align}
M=Q(-r)\,R(\phi)\,Q(r)=\begin{pmatrix}\alpha&\beta\\ \beta^\ast&\alpha^\ast\end{pmatrix},\qquad
\beta\propto i\,\sin\phi.
\end{align}
The $\beta$ coefficient, which (up to passive rotations) sets the squeezing phase $\vartheta_k$, acquires a controllable imaginary part via the dwell phase $\phi$. By implementing two distinct sequences $(r_1,\phi_1)$ and $(r_2,\phi_2)$ for the two superposed histories $\mathcal{T}_{1,2}$, the phase difference $\vartheta_k^{(1)}-\vartheta_k^{(2)}$ (and thus $\arg z_k$) becomes a tunable experimental knob for controlling the TMJS interference.

This brings us to the final answer for our motivating question. The appropriate "vacuum" for an inertial observer is the \emph{in-vacuum on $\mathscr{I}^-$}. The accelerating mirror \emph{changes the quantum state} via an $SU(1,1)$ pair-creation process. A single mirror history (standard DCE) produces a pure \emph{Gaussian} TMSS. In a Gaussian state, all properties are determined by the first and second moments (Wick's theorem), and all connected cumulants beyond second order are zero.

By contrast, the TMJS from superposed histories, Eq.~\eqref{eq:tmjs_mirror_construct}, is fundamentally \emph{non-Gaussian} (as it is a superposition of two distinct Gaussian states). This non-Gaussianity has a direct, operational signature: the intensity-based coherences $g^{(k)}$ are no longer fixed by the second moments. Instead, the interference terms controlled by $\delta$ render the $g^{(k)}$ \emph{phase-steerable}, providing a clear experimental signature that is absent in any single-history, Gaussian DCE.

\section{Conclusion}
\label{sec:conclusion}

In this work, we have introduced and analytically solved the TMJS, a non-Gaussian state defined as a coherent superposition of two distinct two-mode squeezed states. Our central goal was to synthesize the controllable, interference-based non-Gaussianity of the Janus program with the fundamental thermal structure of quantum field theory in curved spacetime, as exemplified by the Unruh effect.

We derived a complete analytical framework for the TMJS photon statistics, applicable to arbitrary $k$-th order correlations. The cornerstone is the family of squeezing polynomials $P_k(x)$ (Eq.~\ref{eq:Fk_final_poly}) governing the diagonal kernels. Extending to the off-diagonal (cross-state) kernels, we obtained closed-form master expressions for the single-mode marginal coherences $g_a^{(k)}(\Psi)$ (Eq.~\ref{eq:tmjs_gka_def}) and the cross-mode coherences $g_{ab}^{(k)}(\Psi)$ (Eq.~\ref{eq:tmjs_gkab_def}).

Our visualization of the coherence landscapes in Sec.~\ref{sec:vis} reveals the consequences of this construction. First, the baseline behaviors are established: the cross-mode coherences $g_{ab}^{(k)}$ of a single TMSS diverge as $r\!\to\!0$ (Fig.~\ref{fig:fig_pk_polynomials}), whereas the single-mode marginals are thermal, $g_a^{(k)}=k!$, in the symmetric setting ($r=s$) with $\Delta=0$ (see the nearly flat top-left panels of Figs.~\ref{fig:single_mode_g2_landscape}–\ref{fig:single_mode_g4_landscape}). Thus, when the squeezing phases are aligned the TMJS reproduces the Unruh-thermal statistics of the marginal.

Introducing a non-trivial relative squeezing phase ($\Delta\neq 0$) activates the Janus phase $\delta$ as a master control knob. Near the antisymmetric point $(\Delta,\delta)=(\pi,\pi)$, destructive interference \emph{strongly suppresses} higher-order correlations: the $r\!\to\!0$ divergence of $g_{ab}^{(k)}$ is inverted into a many-orders-of-magnitude reduction (Fig.~\ref{fig:janus_coherence_switch}), and the single-mode coherences $g_a^{(k)}$ become deeply sub-Poissonian (the dark valleys in Figs.~\ref{fig:single_mode_g2_landscape}–\ref{fig:single_mode_g4_landscape}). Importantly, this suppression is not a trivial intensity fade-out: the mean photon number remains finite and exhibits a pronounced \emph{trough} near the same phase setting (Fig.~\ref{fig:brightness_landscape}), evidencing genuine phase-coherent cancellation rather than depletion.

Conceptually, the TMJS provides a non-Gaussian generalization of the thermofield double (TFD) state. While the standard Unruh effect supplies a \emph{static} Gaussian resource (the Minkowski vacuum as a TMSS), our construction bridges to \emph{dynamic}, state-generating processes. As discussed in Sec.~\ref{sec:physical_realization}, the TMJS arises naturally from coherent superpositions of distinct Dynamical Casimir-effect (“accelerating mirror”) trajectories, unifying observer-dependent and dynamically generated particle-pair states within a single interference-aware framework.

These results establish the TMJS as a versatile, interference-enhanced reference state for relativistic quantum information. It offers a tunable handle for engineering Unruh–DeWitt \cite{Unruh1976, Einstein100, Colosi2009Rovelli} detector responses and for optimizing protocols that leverage higher-order statistics. In particular, phase-steerable correlation suppression supports coherent-cancellation strategies to mitigate noise or selectively enhance signals in accelerated quantum technologies. Future directions include quantifying TMJS entanglement and non-Gaussianity measures, extending to multipartite relativistic settings, and exploiting the phase-contrast of higher moments for non-Gaussian quantum metrology in curved spacetime.

\appendix

\section{Proof of the SU(1,1) Disentangling Theorem}
\label{app:su11_proof}

We provide a proof for the disentangling identity, Eq. \eqref{eq:su11_disentangle}.
The operator is $S_2(\xi) = \exp\{X\}$, where $X = \xi K_+ - \xi^* K_-$. Let $\xi = r e^{i\theta}$.
The generators satisfy $[K_0, K_\pm] = \pm K_\pm$ and $[K_+, K_-] = -2K_0$.

We seek to find the functions $A, B, C$ such that
\begin{align}
S_2(\xi) = \exp\{A K_+\} \exp\{B K_0\} \exp\{C K_-\}.
\label{eq:su11_ansatz}
\end{align}
This is a standard operator ordering problem. A common method is to solve the Wei-Norman differential equations. Let $S(\lambda) = \exp\{\lambda X\}$ and assume a disentangled form $S(\lambda) = \exp\{A(\lambda) K_+\} \exp\{B(\lambda) K_0\} \exp\{C(\lambda) K_-\}$. The goal is to find $A(1), B(1), C(1)$.

We differentiate the ansatz \eqref{eq:su11_ansatz} with respect to $\lambda$:
\begin{align}
\frac{dS}{d\lambda} = \left(\frac{dA}{d\lambda} K_+\right) S + e^{A K_+} \left(\frac{dB}{d\lambda} K_0\right) e^{-A K_+} S + e^{A K_+} e^{B K_0} \left(\frac{dC}{d\lambda} K_-\right) e^{-B K_0} e^{-A K_+} S.
\label{eq:su11_deriv1}
\end{align}
We use the operator identities (Hadamard's lemma):
$e^{X} Y e^{-X} = Y + [X,Y] + \frac{1}{2!}[X,[X,Y]] + \dots$
\begin{align}
e^{A K_+} K_0 e^{-A K_+} &= K_0 - [A K_+, K_0] = K_0 - A K_+ \\
e^{B K_0} K_- e^{-B K_0} &= K_- e^{-B} \\
e^{A K_+} K_- e^{-A K_+} &= K_- - 2 A K_0 + A^2 K_+
\end{align}
Substituting these into \eqref{eq:su11_deriv1} gives:
\begin{align}
\frac{dS}{d\lambda} S^{-1} = \frac{dA}{d\lambda} K_+ + \frac{dB}{d\lambda} (K_0 - A K_+) + \frac{dC}{d\lambda} e^{-B} (K_- - 2 A K_0 + A^2 K_+)
\label{eq:su11_deriv2}
\end{align}
We also know that $\frac{dS}{d\lambda} S^{-1} = X = \xi K_+ - \xi^* K_-$.
By equating the coefficients of $K_+, K_0, K_-$ in \eqref{eq:su11_deriv2}, we get a system of coupled differential equations:
\begin{align}
\text{coeff}(K_+): \quad \xi &= \frac{dA}{d\lambda} - A \frac{dB}{d\lambda} + A^2 e^{-B} \frac{dC}{d\lambda} \\
\text{coeff}(K_0): \quad 0 &= \frac{dB}{d\lambda} - 2 A e^{-B} \frac{dC}{d\lambda} \\
\text{coeff}(K_-): \quad -\xi^* &= e^{-B} \frac{dC}{d\lambda}
\end{align}
We solve this system with initial conditions $A(0)=B(0)=C(0)=0$ (since $S(0)=I$).
From the third equation: $\frac{dC}{d\lambda} = -\xi^* e^{B}$.
Substitute this into the second: $\frac{dB}{d\lambda} = 2 A e^{-B} (-\xi^* e^{B}) = -2 A \xi^*$.
Substitute both into the first: $\xi = \frac{dA}{d\lambda} - A (-2 A \xi^*) + A^2 e^{-B} (-\xi^* e^{B}) = \frac{dA}{d\lambda} + 2 A^2 \xi^* - A^2 \xi^* = \frac{dA}{d\lambda} + A^2 \xi^*$.
So we have the system:
\begin{align}
\frac{dA}{d\lambda} &= \xi - A^2 \xi^* \label{eq:riccati} \\
\frac{dB}{d\lambda} &= -2 A \xi^* \\
\frac{dC}{d\lambda} &= -\xi^* e^{B}
\end{align}
Equation \eqref{eq:riccati} is a Riccati equation for $A$. We can verify the solution $A(\lambda) = \frac{\xi}{|\xi|} \tanh(|\xi| \lambda) = \frac{\xi}{r} \tanh(r \lambda)$.
Let's check: $\frac{dA}{d\lambda} = \frac{\xi}{r} (r \text{sech}^2(r \lambda)) = \xi (1 - \tanh^2(r \lambda))$.
RHS of \eqref{eq:riccati} is: $\xi - (\frac{\xi}{r} \tanh(r \lambda))^2 \xi^* = \xi - \frac{\xi^2 \xi^*}{r^2} \tanh^2(r \lambda) = \xi - \frac{\xi (r^2)}{r^2} \tanh^2(r \lambda) = \xi (1 - \tanh^2(r \lambda))$.
The solution is correct.

Now we solve for $B(\lambda)$:
$\frac{dB}{d\lambda} = -2 \left(\frac{\xi}{r} \tanh(r \lambda)\right) \xi^* = -2 \frac{r^2}{r} \tanh(r \lambda) = -2r \tanh(r \lambda)$.
$B(\lambda) = \int_0^\lambda -2r \tanh(r \lambda') d\lambda' = -2 \ln(\cosh(r \lambda')) \Big|_0^\lambda = -2 \ln(\cosh(r \lambda))$.
So, $e^{B(\lambda)} = \text{sech}^2(r \lambda) = 1 - \tanh^2(r \lambda)$.

Finally, we solve for $C(\lambda)$:
$\frac{dC}{d\lambda} = -\xi^* e^{B(\lambda)} = -\xi^* \text{sech}^2(r \lambda)$.
$C(\lambda) = \int_0^\lambda -\xi^* \text{sech}^2(r \lambda') d\lambda' = -\xi^* \left( \frac{1}{r} \tanh(r \lambda') \right) \Big|_0^\lambda = -\frac{\xi^*}{r} \tanh(r \lambda)$.

At the end of the evolution, $\lambda=1$:
\begin{align}
A(1) &= \frac{\xi}{r} \tanh(r) = e^{i\theta} \tanh(r) \equiv \alpha \\
B(1) &= -2 \ln(\cosh r) = \ln(\text{sech}^2 r) = \ln(1 - \tanh^2 r) = \ln(1 - |\alpha|^2) \\
C(1) &= -\frac{\xi^*}{r} \tanh(r) = -e^{-i\theta} \tanh(r) = -\alpha^*
\end{align}
Substituting these $A(1), B(1), C(1)$ back into the ansatz \eqref{eq:su11_ansatz} gives:
\begin{align}
S_2(\xi)
= \exp\big\{\alpha K_+\big\}\, \exp\big\{\ln(1 - |\alpha|^2) K_0\big\}\, \exp\big\{-\alpha^* K_-\big\},
\end{align}
which exponentiates to the final form:
\begin{align}
S_2(\xi)
= \exp\big\{\alpha K_+\big\}\, (1-|\alpha|^2)^{K_0}\, \exp\big\{-\alpha^* K_-\big\}.
\end{align}
This completes the proof.

\bibliographystyle{jhep}
\bibliography{Unruh+SqueezedRef}

\providecommand{\href}[2]{#2}\begingroup\raggedright\begin{thebibliography}{10}

\bibitem{Israel1976}
W.~Israel, \emph{{Thermo-field dynamics of black holes}}, \href{https://doi.org/https://doi.org/10.1016/0375-9601(76)90178-X}{\emph{Physics Letters A} {\bfseries 57} (1976) 107}.

\bibitem{Unruh1976}
W.G.~Unruh, \emph{{Notes on black-hole evaporation}}, \href{https://doi.org/10.1103/PhysRevD.14.870}{\emph{Phys. Rev. D} {\bfseries 14} (1976) 870}.

\bibitem{Hawking1975}
S.W.~Hawking, \emph{{Particle Creation by Black Holes}}, \href{https://doi.org/10.1007/BF02345020}{\emph{Commun. Math. Phys.} {\bfseries 43} (1975) 199}.

\bibitem{Davies1975}
P.C.W.~Davies, \emph{{Scalar production in Schwarzschild and Rindler metrics}}, \href{https://doi.org/10.1088/0305-4470/8/4/022}{\emph{Journal of Physics A: Mathematical and General} {\bfseries 8} (1975) 609}.

\bibitem{Fulling1973}
S.A.~Fulling, \emph{{Nonuniqueness of Canonical Field Quantization in Riemannian Space-Time}}, \href{https://doi.org/10.1103/PhysRevD.7.2850}{\emph{Phys. Rev. D} {\bfseries 7} (1973) 2850}.

\bibitem{DeWitt1975PhysicsRep}
B.S.~DeWitt, \emph{{Quantum field theory in curved spacetime}}, \href{https://doi.org/https://doi.org/10.1016/0370-1573(75)90051-4}{\emph{Physics Reports} {\bfseries 19} (1975) 295}.

\bibitem{Birrell_Davies1982}
N.D.~Birrell and P.C.W.~Davies, \emph{{Quantum Fields in Curved Space}}, Cambridge Monographs on Mathematical Physics, Cambridge University Press, Cambridge, UK (1982), \href{https://doi.org/10.1017/CBO9780511622632}{10.1017/CBO9780511622632}.

\bibitem{UnruhWald1984}
W.G.~Unruh and R.M.~Wald, \emph{{What happens when an accelerating observer detects a Rindler particle}}, \href{https://doi.org/10.1103/PhysRevD.29.1047}{\emph{Phys. Rev. D} {\bfseries 29} (1984) 1047}.

\bibitem{Rindler66}
W.~Rindler, \emph{{Kruskal Space and the Uniformly Accelerated Frame}}, \href{https://doi.org/10.1119/1.1972547}{\emph{Am. J. Phys.} {\bfseries 34} (1966) 1174}.

\bibitem{Weedbrook2012gaussian}
C.~Weedbrook, S.~Pirandola, R.~Garc\'{\i}a-Patr\'on, N.J.~Cerf, T.C.~Ralph, J.H.~Shapiro et~al., \emph{{Gaussian quantum information}}, \href{https://doi.org/10.1103/RevModPhys.84.621}{\emph{Rev. Mod. Phys.} {\bfseries 84} (2012) 621}.

\bibitem{Braunstein2005quantum}
S.L.~Braunstein and P.~van Loock, \emph{{Quantum information with continuous variables}}, \href{https://doi.org/10.1103/RevModPhys.77.513}{\emph{Rev. Mod. Phys.} {\bfseries 77} (2005) 513}.

\bibitem{Adesso2014continuous}
G.~Adesso, S.~Ragy and A.R.~Lee, \emph{{Continuous Variable Quantum Information: Gaussian States and Beyond}}, \href{https://doi.org/10.1142/S1230161214400010}{\emph{Open Systems \& Information Dynamics} {\bfseries 21} (2014) 1440001}.

\bibitem{Wu1986generation}
L.-A.~Wu, H.J.~Kimble, J.L.~Hall and H.~Wu, \emph{{Generation of Squeezed States by Parametric Down Conversion}}, \href{https://doi.org/10.1103/PhysRevLett.57.2520}{\emph{Phys. Rev. Lett.} {\bfseries 57} (1986) 2520}.

\bibitem{slusher1985observatio}
R.E.~Slusher, L.W.~Hollberg, B.~Yurke, J.C.~Mertz and J.F.~Valley, \emph{{Observation of Squeezed States Generated by Four-Wave Mixing in an Optical Cavity}}, \href{https://doi.org/10.1103/PhysRevLett.55.2409}{\emph{Phys. Rev. Lett.} {\bfseries 55} (1985) 2409}.

\bibitem{Menicucci2006}
N.C.~Menicucci, P.~van Loock, M.~Gu, C.~Weedbrook, T.C.~Ralph and M.A.~Nielsen, \emph{Universal quantum computation with continuous-variable cluster states}, \href{https://doi.org/10.1103/PhysRevLett.97.110501}{\emph{Phys. Rev. Lett.} {\bfseries 97} (2006) 110501}.

\bibitem{Ohliger2010limitations}
M.~Ohliger, K.~Kieling and J.~Eisert, \emph{Limitations of quantum computing with gaussian cluster states}, \href{https://doi.org/10.1103/PhysRevA.82.042336}{\emph{Phys. Rev. A} {\bfseries 82} (2010) 042336}.

\bibitem{Mari2012Wigner}
A.~Mari and J.~Eisert, \emph{{Positive Wigner Functions Render Classical Simulation of Quantum Computation Efficient}}, \href{https://doi.org/10.1103/PhysRevLett.109.230503}{\emph{Phys. Rev. Lett.} {\bfseries 109} (2012) 230503}.

\bibitem{Albarelli2018Resource}
F.~Albarelli, M.G.~Genoni, M.G.A.~Paris and A.~Ferraro, \emph{{Resource theory of quantum non-Gaussianity and Wigner negativity}}, \href{https://doi.org/10.1103/PhysRevA.98.052350}{\emph{Phys. Rev. A} {\bfseries 98} (2018) 052350}.

\bibitem{Azizi2025Janus}
A.~Azizi, \emph{{The Janus State: A Universal Lower Bound for Second-Order Coherence}},  \href{https://arxiv.org/abs/2506.06397}{{\ttfamily 2506.06397}}.

\bibitem{Azizi2025Janus_higher}
A.~Azizi, \emph{Quantum complementarity ad infinitum: Switching higher-order coherence from infinity to zero},  \href{https://arxiv.org/abs/2507.15890}{{\ttfamily 2507.15890}}.

\bibitem{Azizi2025displacedJanus}
A.~Azizi, \emph{{Tunable Non-Gaussianity and Exact Higher-Order Coherences for Quantum Advantage}}, \href{https://doi.org/10.1103/ggg4-scz9}{\emph{Phys. Rev. A (Accepted)} (2025) } [\href{https://arxiv.org/abs/2508.09234}{{\ttfamily 2508.09234}}].

\bibitem{Kenfack2004}
A.~Kenfack and K.~Życzkowski, \emph{{Negativity of the Wigner function as an indicator of non-classicality}}, \href{https://doi.org/10.1088/1464-4266/6/10/003}{\emph{Journal of Optics B: Quantum and Semiclassical Optics} {\bfseries 6} (2004) 396}.

\bibitem{Moore1970}
G.T.~Moore, \emph{{Quantum Theory of the Electromagnetic Field in a Variable‐Length One‐Dimensional Cavity}}, \href{https://doi.org/10.1063/1.1665432}{\emph{Journal of Mathematical Physics} {\bfseries 11} (1970) 2679}.

\bibitem{Fulling_Davies1976}
S.A.~Fulling and P.C.W.~Davies, \emph{{Radiation from a moving mirror in two dimensional space-time: conformal anomaly}}, \href{https://doi.org/10.1098/rspa.1976.0045}{\emph{Proceedings of the Royal Society of London. A. Mathematical and Physical Sciences} {\bfseries 348} (1976) 393}.

\bibitem{Nation2012Nori}
P.D.~Nation, J.R.~Johansson, M.P.~Blencowe and F.~Nori, \emph{{Colloquium: Stimulating uncertainty: Amplifying the quantum vacuum with superconducting circuits}}, \href{https://doi.org/10.1103/RevModPhys.84.1}{\emph{Rev. Mod. Phys.} {\bfseries 84} (2012) 1}.

\bibitem{Azizi2025Unitary_TFD}
A.~Azizi, \emph{{Unitary reformulation of the thermofield double state and limits of cyclic multimode squeezing}}, \href{https://doi.org/10.1103/9st7-fxhf}{\emph{Phys. Rev. D} {\bfseries 112} (2025) 065008}.

\bibitem{Azizi2023JHEP}
A.~Azizi, \emph{{Kappa vacua: enhancing the Unruh temperature}}, \href{https://doi.org/10.1007/JHEP07(2023)064}{\emph{Journal of High Energy Physics} {\bfseries 2023} (2023) 64}.

\bibitem{Einstein100}
B.S.~DeWitt, \emph{{General Relativity}: {An Einstein Centenary Survey}}, Univ. Pr., Cambridge, UK (1979).

\bibitem{Colosi2009Rovelli}
D.~Colosi and C.~Rovelli, \emph{{What is a particle?}}, \href{https://doi.org/10.1088/0264-9381/26/2/025002}{\emph{Classical and Quantum Gravity} {\bfseries 26} (2008) 025002}.

\end{thebibliography}\endgroup
\end{document}